\documentclass[final,1p,times,authoryear]{elsarticle}
\newcommand{\hide}[1]{}

\usepackage{graphicx}
\usepackage{subfigure}
\usepackage{subfigmat}

\usepackage{amssymb}

\journal{Icarus}

\begin{document}

\begin{frontmatter}



\title{Is the Grand Tack model compatible with the orbital distribution of main belt asteroids?}


\author[inpe]{Rogerio Deienno}
\address[inpe]{Instituto Nacional de Pesquisas Espaciais, Avenida dos Astronautas 1758, CEP 12227-010 S\~ao Jos\'e dos Campos, SP, Brazil}
\ead{rogerio.deienno@gmail.com}

\author[on]{Rodney S. Gomes}
\address[on]{Observat\'orio Nacional, Rua General Jos\'e Cristino 77, CEP 20921-400 Rio de Janeiro, RJ, Brazil}

\author[swri]{Kevin J. Walsh}
\address[swri]{Department of Space Studies, Southwest Space Research Institute, 1050 Walnut St., Boulder, CO 80302, USA}

\author[oca]{Alessandro Morbidelli}
\address[oca]{Laboratoire Lagrange, UMR7293, Universit\'e C\^ote d'Azur, CNRS, Observatoire de la C\^ote d'Azur, Boulevard de l'Observatoire, 06304 Nice Cedex 4, France}


\author[swri]{David Nesvorn\'y}

%

\begin{abstract}

The Asteroid Belt is characterized by the radial mixing of bodies with
different physical properties, a very low mass compared to Minimum Mass
Solar Nebula expectations and has an excited orbital distribution, with
eccentricities and inclinations covering the entire range of values
allowed by the constraints of dynamical stability.  Models of the
evolution of the Asteroid Belt show that the origin of its structure
is strongly linked to the process of terrestrial planet formation. The
Grand Tack model presents a possible solution to the conundrum of
reconciling the small mass of Mars with the properties of the Asteroid
Belt, including the mass depletion, radial mixing and orbital
excitation.  However, while the inclination distribution produced in
the Grand Tack model is in good agreement with the one observed, the
eccentricity distribution is skewed towards values larger than those
found today. Here, we evaluate the evolution of the orbital properties
of the Asteroid Belt from the end of the Grand Tack model (at the end
of the gas nebula phase when planets emerge from the dispersing gas
disk), throughout the subsequent evolution of the Solar System
including an instability of the Giant Planets approximately 400 My
later. Before the instability, the terrestrial planets were modeled on
dynamically cold orbits with Jupiter and Saturn locked in a 3:2 mean
motion resonance.  The model continues for an additional 4.1 Gy after
the giant planet instability. Our results show that the eccentricity
distribution obtained in the Grand Tack model evolves towards one very
similar to that currently observed, and the semimajor axis distribution does the same. The inclination distribution remains
nearly unchanged with a slight preference for depletion at low inclination; this leads to the conclusion that the inclination distribution at the end of the Grand Tack is a bit over-excited. Also, we constrain the primordial eccentricities of
Jupiter and Saturn, which have a major influence on the dynamical
evolution of the Asteroid Belt and its final orbital structure.

\end{abstract}


\end{frontmatter}



\section{Introduction} \label{intro}

The Asteroid Belt is challenging to understand but is critical for
studies of the formation and early evolution of the Solar System. The
orbital configuration of the Asteroid Belt is believed to have been
established in two phases. The first phase dates back to the first
few million years of Solar System's formation and should be studied in
conjunction with the formation of the inner and outer planets,
especially Jupiter and Saturn. The second phase occurred when the
Asteroid Belt witnessed a Giant Planet instability, long after the
damping effects of the gaseous Solar Nebula had dissipated
 
In general, simulations of the dynamical re-shaping of the Asteroid
Belt are made in conjunction with the formation of the inner
planets. The first simulations of terrestrial planet formation \citep{Chambers98} included a set of
planetary embryos uniformly distributed in the inner region of the
Solar System with orbits initially dynamically cold (low eccentricity
and inclination). Through numerical integrations of the equations of
motion of these embryos, adding a model of accretion by collisions,
the system evolves to form planets in the inner region of the Solar
System on stable orbits. While early results about the formation of
terrestrial planets were promising, one of the problems found in these
integrations was related with the final eccentricities of the planets,
which were systematically larger than the real ones. The models
produced more promising results when the presence of a substantial
population of planetesimals was also accounted for; in fact, the
dynamical friction exerted by the planetesimals acted to decrease the
excitation of the planet's final orbits \citep{Chambers01, OBrien06}.

An important ingredient was the presence of Jupiter, which should have completed its formation
much earlier than the inner
planets \citep{Chambers98, Chambers01, Petit01}. Primarily, the
influence of Jupiter on the Asteroid Belt is to promote
destructive collisions (fragmentation) rather than constructive
collisions (accretion) \citep{Petit02}. However, Jupiter alone can not
excite the eccentricity of planetesimals so much as to explain the
current excited orbits of asteroids \citep{Petit02}.  In addition, there
is significant diversity in the physical properties of asteroids found
in the Main Asteroid Belt, but their main taxonomic classes are found
in roughly overlapping distributions -- although S-class bodies
predominate in the inner regions and C-class bodies in the outer
regions \citep[see][]{DeMeo14}. The solution of these issues have
been attributed to the original presence of planetary embryos in the
Asteroid Belt \citep{Petit99,OBrien07}. These embryos, once excited by
Jupiter, would have scattered the orbits of the planetesimals. In the
end, the Asteroid Belt would have been depleted of planetesimals and
totally devoid of embryos.

Despite the many successes in the modeling of the terrestrial planets
and Asteroid Belt by the simulations described above, systematic
problems persisted. The planet formed in the approximate region of
Mars systematically showed a much larger mass than the real Mars \citep[see][]{Raymond09}. An experiment by \citet{Hansen09} found that if
there is sharp outer edge in the initial mass distribution of solids
at about 1.0 AU, then the models consistently reproduce the mass of
Mars.

\citet{Walsh11} proposed a mechanism to modify the original mass
distribution of solids and produce the truncated disk explored by
\citet{Hansen09}, by accounting for the early migration of Jupiter and
Saturn when they were still embedded in the gaseous proto-planetary
disk.  An outcome found in many hydrodynamical models
\citep{Masset01,Morbidelli07,Pierens08,Pierens11,DAngelo12} of the
interaction between giant planets and gaseous disks is that the
type-II inward migration of a Jupiter-mass planet is halted and even
reversed when a second, less massive planet, is formed external to the
first one. This provides the explanation for why Jupiter did not
migrate to very close the Sun, as is seen for giant planets in many
other planetary systems \citep{Udry07,Cumming08}. Instead, Jupiter
would have migrated first inwards, then outwards. Because of the
change in direction of the orbital motion of Jupiter (a ``tack'' in
sailor's jargon), the \citet{Walsh11} model is named the ``Grand
Tack''. The timing of the formation of Saturn is constrained by
  the mass distribution of the terrestrial planets, which are best
  reproduced when Jupiter reverses migration at 1.5 AU and truncates
  the disk at 1 AU.

The migration of Jupiter would have strongly affected any
planetesimals formed in the present-day Asteroid Belt, with a primary
consequence of substantially depleting the entire region of small
bodies.  The inward migration phase primarily pushes the asteroids
originally inside of Jupiter's orbit (named ``S-class'' in
\citet{Walsh11}) down to lower semimajor axes (inside of 1~AU), though
\citet{Walsh11} found that about 10\% of these bodies are scattered
outward onto orbits with semimajor axis $a$ between 4-10 AU. During the
outward migration of Jupiter and Saturn, these bodies are encountered
again, and about 1\% are scattered back into the Asteroid
Belt. Meanwhile Jupiter and Saturn eventually encounter primitive
planetesimals (titled ``C-class'' in \citet{Walsh11}), and a fraction
of a percent of these are also scattered into the Asteroid Belt. This
provides, at the time when the gas nebula has dispersed, a final belt
which is depleted in mass by a factor of about 1,000, that contains
two different classes of bodies partially mixed in heliocentric
distance and with orbits excited in eccentricities and inclinations
(although the final eccentricity distribution does not match well the
current one, as discussed below).

Numerous constraints, such as the ages of the last impact basins on
the Moon \citep{Bottke07}, the impact age distribution of HED meteorites \citep{Marchi13},
and the small total chondritic mass accreted by the Moon since its
formation \citep{Morbidelli12}, point to an epoch of increased bombardment in the inner Solar System about $\sim$400$-$700 My after the removal of gas from
the proto-planetary disk (whereas the Grand Tack happened before the
removal of the gas).  This period of increased bombardment is usually called
``Terminal Lunar Cataclysm'' or ``Late Heavy Bombardment'' (LHB) \citep[see][for reviews]{Hartmann00,Chapman07}, and we
will adopt the LHB nomenclature here.  The origin of the
LHB has been linked to a dynamical upheaval in the outer Solar System
frequently referred to as the ``Nice model'' \citep{Gomes05,
  Levison11, Bottke12}. During this dynamical upheaval the giant
planets would have suffered an instability and a period of mutual
close encounters that radically changed their orbits.  In turn, the
orbital change of the giant planets would have severely affected the
distribution of the asteroids in the main belt \citep{Morbidelli10}.
The best guess on when this instability occurred, from various
constraints, is 4.1 Gy ago \citep{Bottke12,Morbidelli12}.

This is important because the final Asteroid Belt in \citet{Walsh11}
lacks objects with small eccentricities. Indeed, according to the
Grand Tack model, the eccentricity distribution expected for the
Asteroid Belt at the time when the gas nebula dispersed, some 3-10 Myr
after the emergence of first solids and roughly 4.5 Gyr ago, peaks
around 0.4. On the other hand, the current distribution of the
Asteroid Belt peaks around 0.1 \citep{Morbidelli2015}. It has never
been studied whether the Grand Tack final distribution could evolve to
one similar to what we see today due to the perturbations caused by
the giant planet instability during the Nice Model. The goal of this
paper is to present such a study.

In this work, we will study in detail the evolution of the Asteroid
Belt orbital structure, from the end of the Grand Tack model to the
giant planet orbital instability, through the instability phase, and
finally during the last $\sim$4 Gy until today.

This work will therefore unfold as follows: section \ref{method}
explains our Solar System's configuration and the method used for the
numerical simulations. In section \ref{results} we discuss the
dynamical evolution of the Asteroid Belt throughout the various phases
of the Solar System's evolution, as well as the influence of the
primordial eccentricities of Jupiter and Saturn on the final structure
of the Asteroid Belt. Finally, section \ref{conclusion} summarizes the
main conclusions of this paper.

A complementary study, approaching the problem from a different
perspective, has been recently presented in \cite{Roig15}. We will
compare our results with theirs at the end of section \ref{post-LHB}.

\section{Methods} \label{method}

\begin{figure*}[t!]
\begin{subfigmatrix}{2}
\includegraphics*{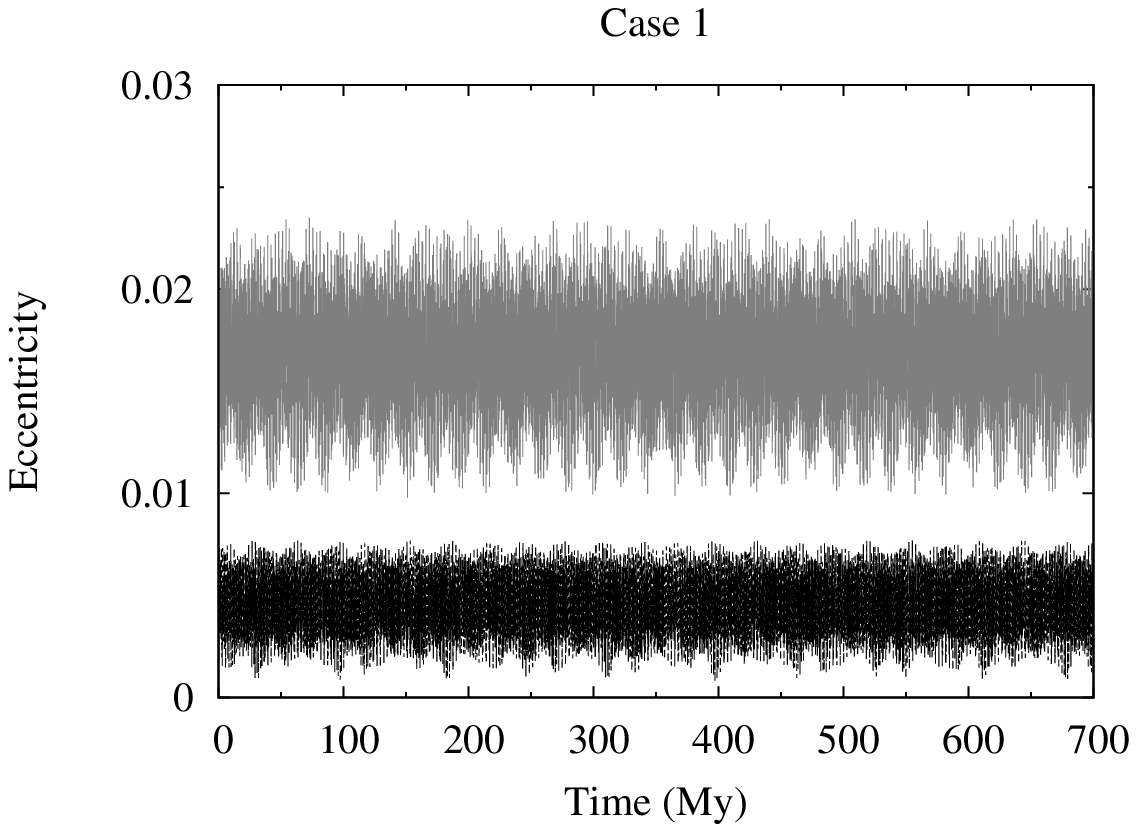}
\includegraphics*{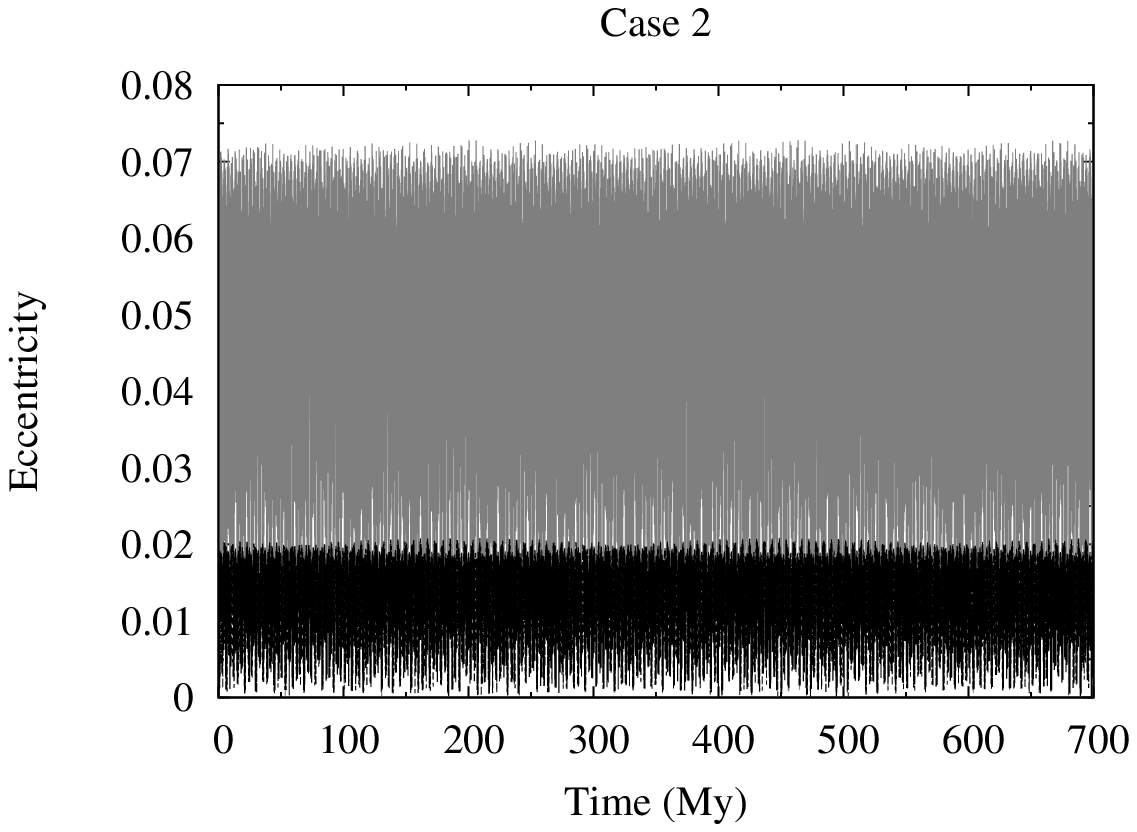}
 \end{subfigmatrix}
\caption{Eccentricity behavior of Jupiter (black) and Saturn (gray) before the planetary instability as a function of time. These two planets are in coplanar orbits, and locked in their 3:2 MMR with Jupiter at 5.4 AU. Left: Case 1 is a initial configuration from \citet{Morbidelli07} where planets evolve into a resonant configuration in a hydrodynamical simulation. Right: Case 2 is an arbitrary case where planets are placed in resonant orbits, allowing large variations in eccentricity. \label{fig1}}
\end{figure*}

Our numerical simulations contain five planets, Venus, Earth, Mars,
Jupiter, and Saturn (all with their current masses), plus 10,000
massless particles representing the final outcome of the Asteroid Belt
in the Grand Tack simulations (described below). 
Uranus, Neptune, and the putative extra ice giant invoked in 
\citet{Nesvorny12} are not included in any of our simulations (the same 
applies for the planet Mercury). These planets are indeed too far from the 
Asteroid Belt to have any important direct effect on its structure. A caveat, 
however, is that in some simulations the extra ice giant is sent 
temporarily onto an Asteroid Belt-crossing orbit before being ultimately 
ejected from the Solar System. In the case the Asteroid Belt-crossing 
phase last for sufficiently long time, the effects of this body on the 
Asteroid Belt should be taken into account \citep[see][]{Brasil16}. 
Nevertheless, because we have limited understanding on how exactly the 
evolution of the planets occurred (i.e. how deep a planet crossed the 
Asteroid Belt and for how long), we prefer to neglect this putative Asteroid 
Belt-crossing event and focus only on the effects induced by Jupiter, 
Saturn and the terrestrial planets, whose evolution is better constrained.

The orbital configuration of the five planets was chosen as follows:

\begin{itemize}

\item {\it Before the planetary instability:}
The terrestrial planets are modeled to be on simple planar orbits
\citep{Brasser13}. Jupiter and Saturn, also on planar orbits, were
locked in their mutual 3:2 MMR with Jupiter at 5.4 AU. Two cases,
differing for the eccentricity of the giant planets
(Fig. \ref{fig1}) are considered. Case 1 (left) is an initial
configuration from the hydrodynamical simulations of
\citet{Morbidelli07}, where the planets migrated into the resonance
and preserved orbits with very small eccentricities due to the
tidal damping effect exerted by the gas disk. Case 2 (right) is an
arbitrary case where the planets are placed in resonant orbits, but
with larger eccentricities (although still smaller than their current
values). For this phase (for both cases), the numerical integrations are continued for
400 My (i.e. until 4.1 Gy ago), the best estimate of the time of the
giant planet instability \citep{Gomes05,Bottke12,Morbidelli12}.

\item {\it After the planetary instability:}
The initial asteroid's orbits are those recorded at the end of the
simulations of the previous phase, but the planets are placed onto
their present-day orbits. By doing this, we model the instantaneous transition of the planets from their pre-instability orbits to the current orbits. In other words, we simulate the response of the Asteroid Belt to the change of the planetary orbits, but we assume that the actual dynamical path that the planets followed in this
transition had no important effects. We take this simple
approach because precisely what happened during the planetary
instability is unknown, and many different paths could have been
followed by the giant planets \citep{Nesvorny12}. Certainly, no
instability simulation can have the pretension to reproduce exactly
what happened. We know from previous studies
\citep{Brasser09,Morbidelli10} that the change of the giant planets'
orbits had to be fast and of the ``jumping-Jupiter'' type,
i.e. dominated by a few impulsive events due to mutual planetary
encounters. Thus, we believe that substituting the original planetary
orbits with the current orbits is the best approximation of the real
planetary evolution we can implement without the risk of incurring
into arbitrary simulation artefacts. In particular, by using the current
planetary system, the locations and strengths of all resonances
exactly match the real ones, whereas no singular instability
simulation would finish with precisely a replica of the current-day
planetary system.  For the same reasons, this approach has already
been adopted by \citet{Bottke12} in their study of the projectile flux
from the Asteroid Belt to the terrestrial planets.  The opposite
approach (i.e. simulating the dynamics of the giant planets during the
instability phase) has been taken by \citet{Roig15}.
The second phase of simulations were conducted for 4.1 Gy, until a
total timescale of 4.5 Gy was covered by the combination of the two
phases of the Solar System evolution.

\end{itemize}

 Concerning the asteroid distribution we consider two cases. In our
  nominal case we use the asteroid distribution resulting from the
  simplest of all Grand Tack scenarios, namely the one that only
  included Jupiter and Saturn (Figure 2. of Supplementary Information
  of \citet{Walsh11}, referenced as ``2-planet Grand Tack
  simulation'' hereafter). However, for comparison we also consider
  the distribution generated in the Grand Tack scenario that also include
  Uranus and Neptune (presented in the main
  paper of \citet{Walsh11}, referenced as ``4-planet Grand Tack
  simulation'' hereafter).
 
Because the original Grand Tack simulations had only $\sim$1,000 particles in the final sample in the Asteroid Belt, in order to improve statistics we generated 10,000 massless particles matching the basic traits of the $a,e,i$ distribution obtained in the 2-planet Grand Tack simulation, with the following properties:

\begin{itemize}
\item Eccentricity distribution is a normal distribution with mean of 0.38 and sigma of 0.17.

\item Inclination distribution is a Rayleigh distribution with a sigma of 10 degrees.

\item Semimajor axis distribution is nearly an uniform distribution spread over the range 1.8--3.6 AU.
\end{itemize} 

While these functional fits reproduce each distribution individually, combined they produced an excess of objects with high-eccentricity and low-inclination compared to the distribution of  asteroids at the end of the Grand Tack simulations.
To remove this excess, the 10,000 particles were re-sampled in bins of width 0.05 in eccentricity and 5 degrees in inclination.
In each bin, we allowed the total number of particles to be at most 4 times more numerous than the asteroids at the end of the Grand Tack simulations in the same bin.  If there were no original Grand Tack asteroids in the bin, a maximum of 4 particles from the functional fits was allowed. This re-sampling limited the excess population of high-eccentricity and low-inclination bodies found in the 10,000 particle distributions, and resulted in a final population of 6,424 bodies. The advantage of this procedure,  relative to using directly the Grand Tack asteroids as initial conditions,  is that the functional fits allow us to generate more particles (but this could have been achieved also by cloning the Grand Tack asteroids)  and also place some particles in bins originally with zero Grand Tack asteroids, thus making the distribution smoother.

For the 4-planet Grand Tack simulation we simply resampled the above distribution attributing to each particle a ``weight'', representing the probability that said particle exists at the end of the 4-planet Grand Tack run. Because the semimajor axis and inclination distributions in the 2-planet and 4-planet  Grand Tack simulations are basically the same, these weights are computed from the eccentricity distributions only, which are significantly different. These weights are then used to build a new final distribution from that obtained in our nominal case.

In all simulations all planets interact with each other while also
perturbing the test particles. Test particles interact with the
planets but not among themselves.  The integrations have been
conducted using Mercury \citep{Chambers99}, in the Hybrid option with
a time step of 10 days (more than 20 steps per orbital period of
Venus).

\section{Results} \label{results}

\begin{figure}[h!]
\includegraphics{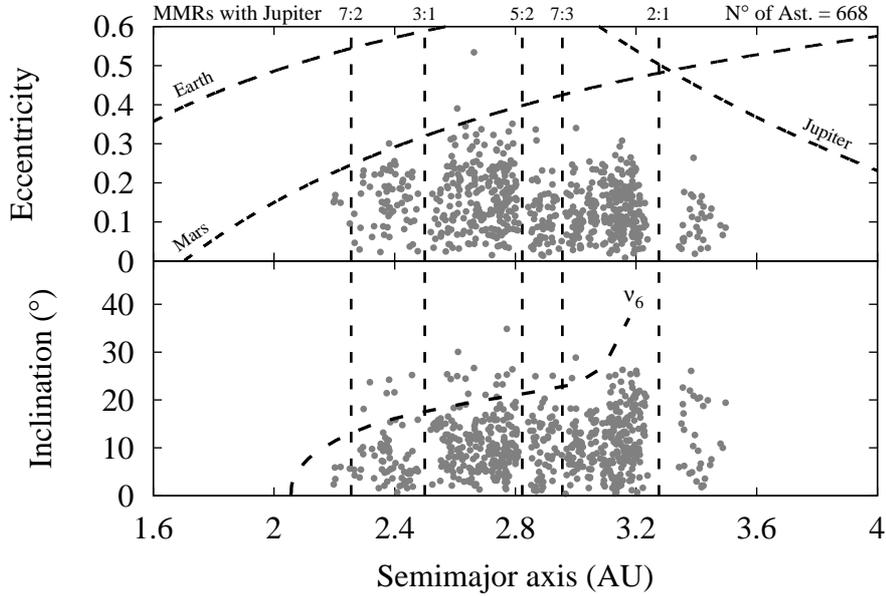}
\caption{Current structure of the Asteroid Belt with all H$<$10 objects from the Minor Planet Center catalog. Gray dots represent the asteroids. Curved dashed lines in the top panel represent the boundaries for Earth-, Mars-, and Jupiter-crossing orbits (from left to right). In the bottom panel, the curved dashed line represents the current location of the $\nu_6$ secular resonance, which occurs when the precession rate of an asteroid's longitude of perihelion is equal to the mean precession rate of the perihelion of Saturn. Vertical dashed lines show MMRs between asteroids and Jupiter. \label{fig2}}
\end{figure}

Figure \ref{fig2} shows the current structure of the Asteroid Belt
plotting all objects with absolute magnitude H$<$10 from the Minor
Planet Center
catalog\footnote{http://www.minorplanetcenter.org/iau/MPCORB.html},
where the Main Asteroid Belt is assumed to be observationally complete
at this size \citep{Jedicke02}. Our goal is to verify if it is
possible that the orbital configuration of the Asteroid Belt at the
end of the Grand Tack model could evolve to one similar to that shown
in this figure. To do so, we considered the two phases of the Solar
System evolution previously discussed in section \ref{method}.

\subsection{Before the planetary instability} \label{pre-LHB}

\begin{figure*}[h!]
\begin{subfigmatrix}{2}
\includegraphics*{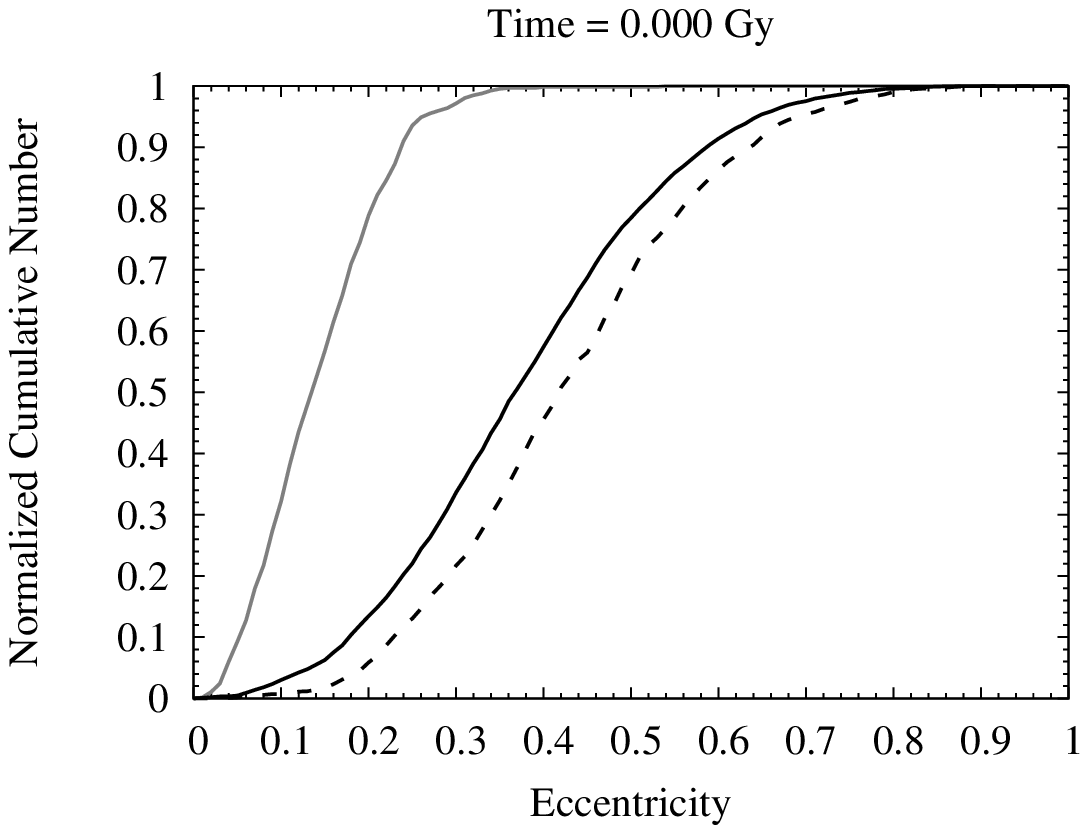}
\includegraphics*{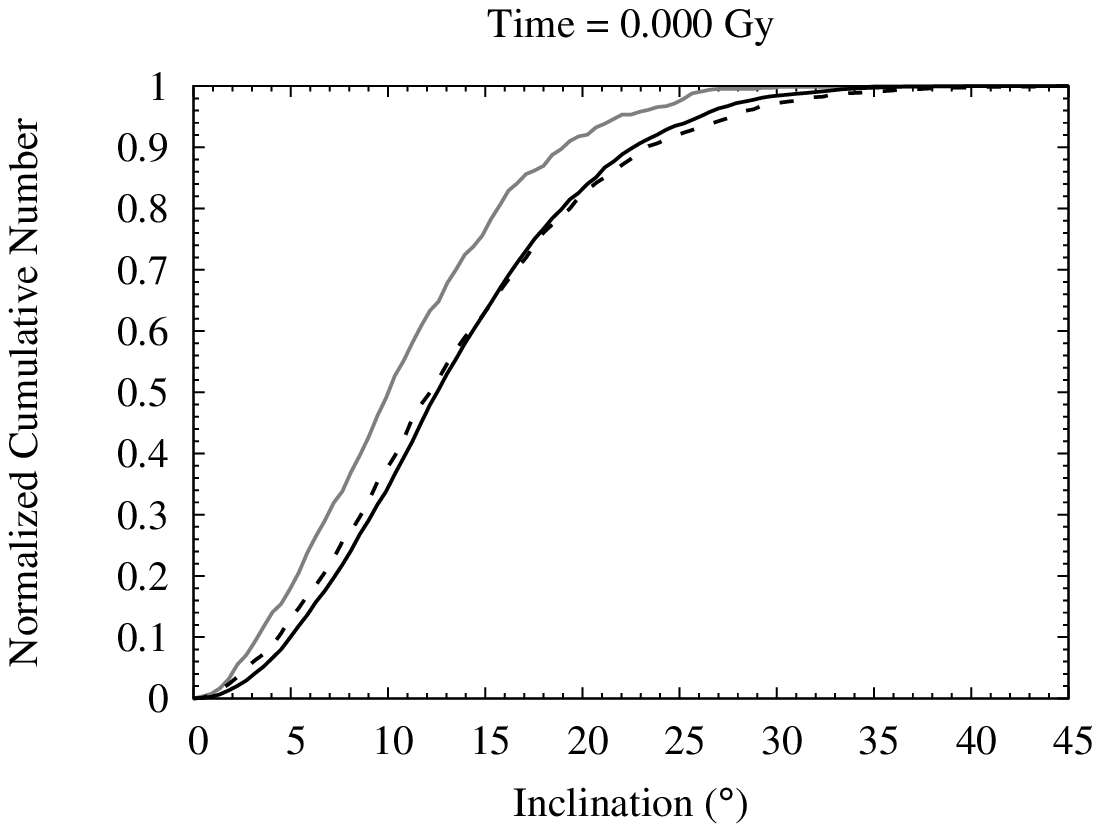}
\includegraphics*{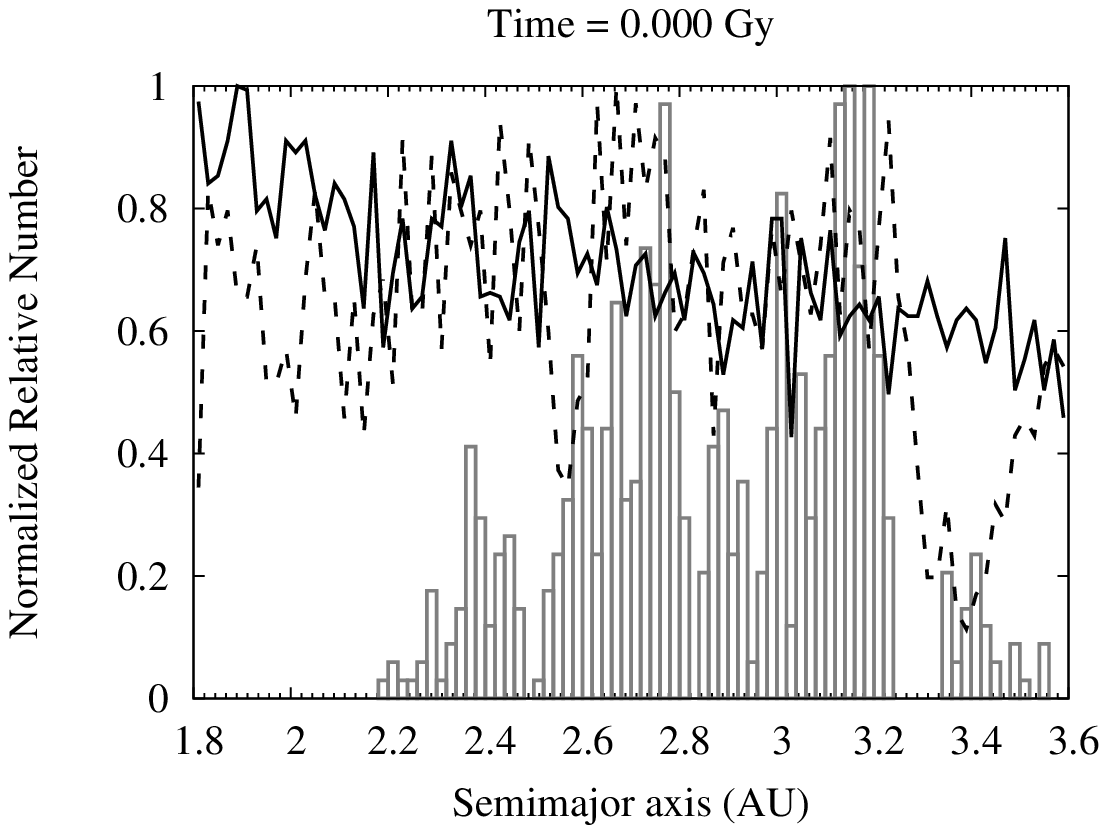}
\includegraphics*{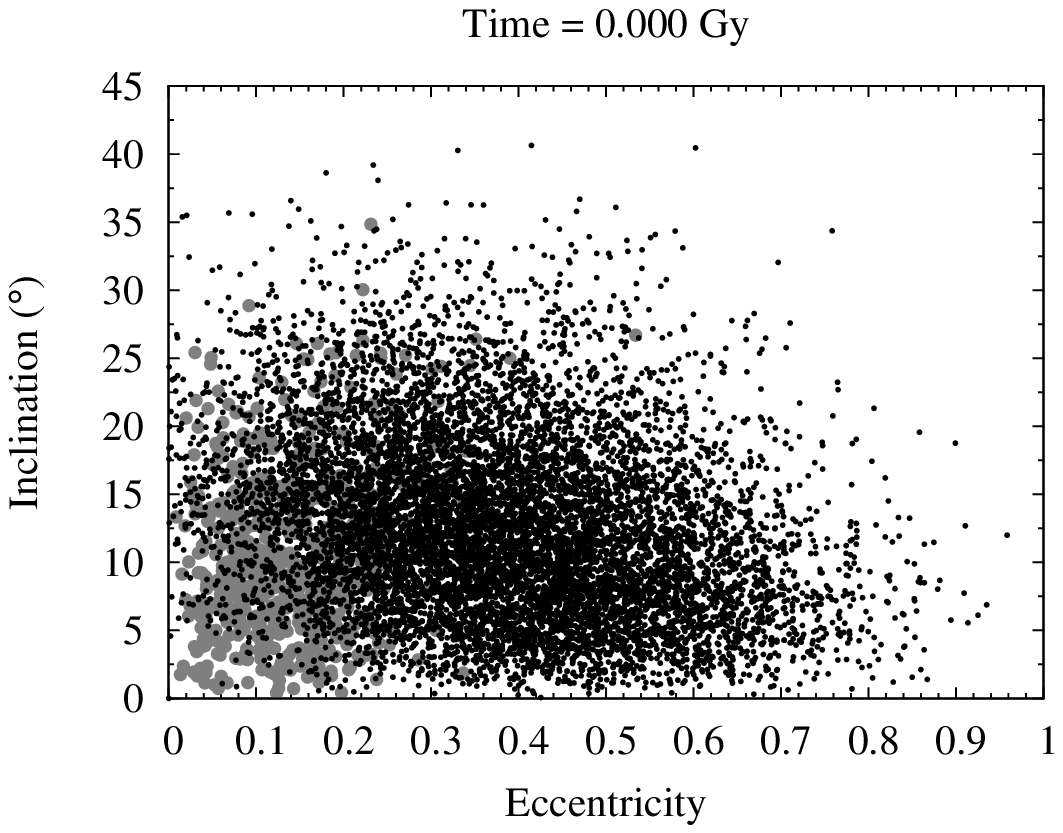}
 \end{subfigmatrix}
\caption{Normalized cumulative number of asteroids as a function of the eccentricity (top left) and inclination (top right). Normalized relative number of asteroids as a function of semimajor axis (bottom left). The width $\Delta a$ of the bins of the histogram is 0.02 AU in semimajor axis. In all these panels the gray line refers to the real distribution, the solid and dashed black lines refer to the final distributions in the 2-planet and 4-planet Grand Tack simulations respectively.
Bottom right: inclination as a function of the eccentricity. Large gray dots represent the real asteroids with H$<$10 from the Minor Planet Center catalog. Small black dots represent our initial distribution generated as described in section 2 (the distribution at the end of the 2-planet Grand Tack simulation, i,e., after the dispersion of the gas nebula). \label{fig3}}
\end{figure*}

Figure \ref{fig3} shows the current Asteroid Belt orbital distribution
(gray line) compared with the initial conditions used in this
experiment that were generated from the Grand Tack simulations (as
described in section \ref{method}); the solid line is for the result
of the 2-planet Grand Tack simulation and the dashed line is for the 4-planet
simulation.  With the exception of the inclination distributions, which match fairly well, the
Grand Tack and the real distributions are clearly different, i.e., the
eccentricity distributions in the Grand Tack simulations are skewed towards large values, and the semimajor axis distributions are
nearly uniform instead of having the clear gaps found in today's
population. The 2-planet and 4-planet Grand Tack simulations also give eccentricity distributions quite different from each other, with the distribution obtained in the 4-planet simulation being even more skewed to large values. Many particles in the final Grand Tack distributions,
however, must clearly be unstable, because their small semimajor axis
and/or large eccentricity make them planet-crossing.

\begin{figure*}[h!]
\begin{subfigmatrix}{2}
\includegraphics*{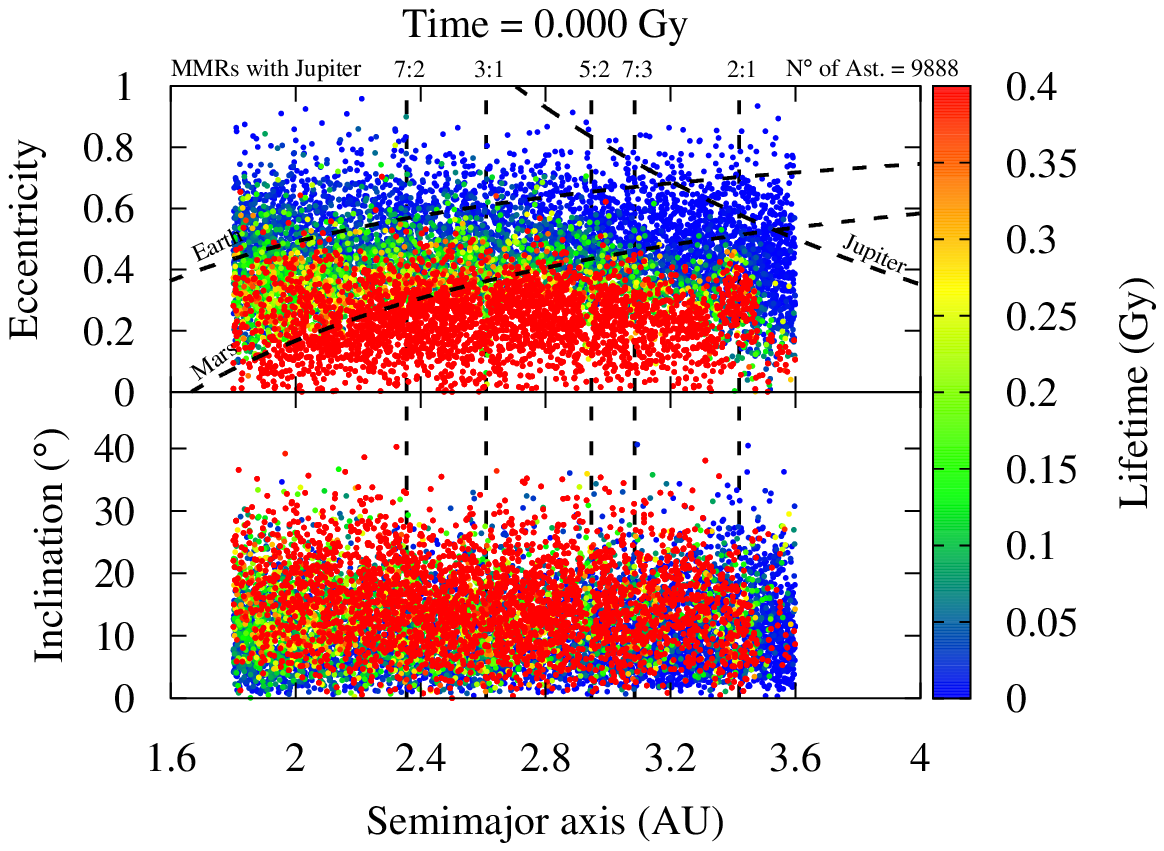}
\includegraphics*{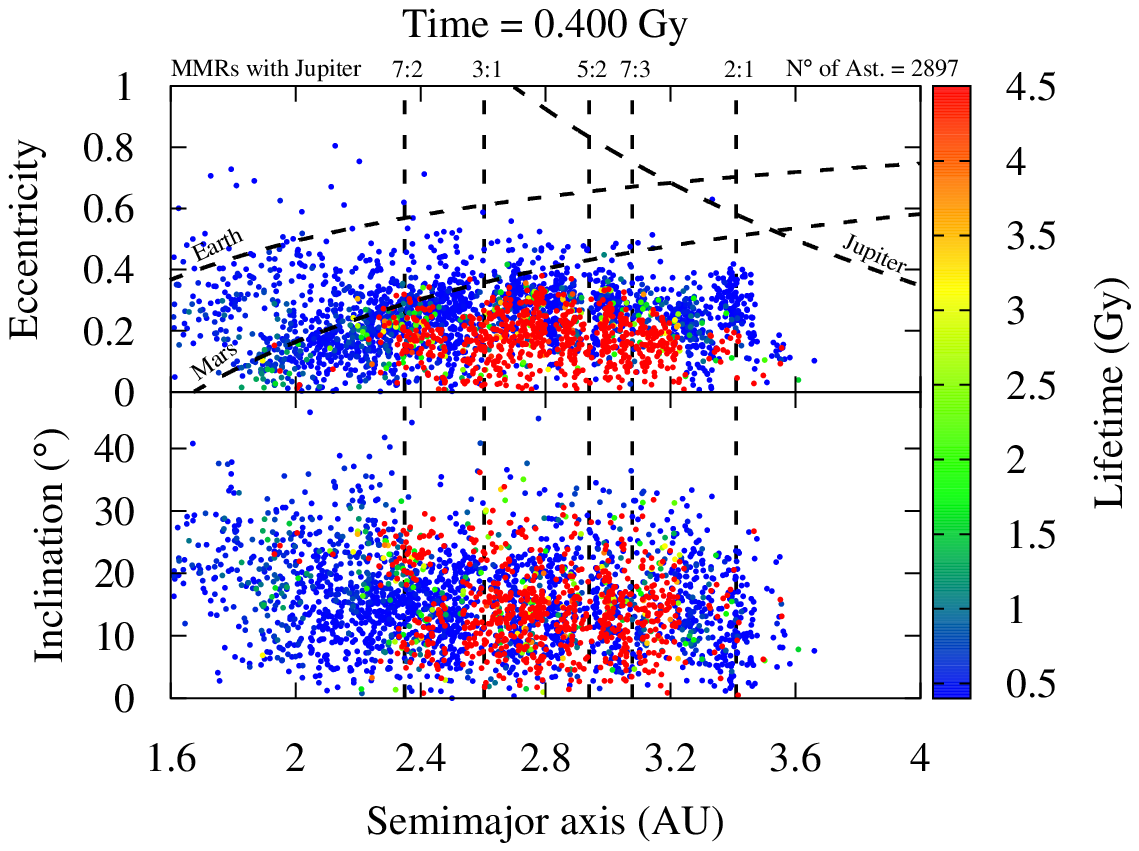}
 \end{subfigmatrix}
\caption{Left: Distribution of the particles resulting from the Grand Tack 2-planet simulation, which consitutes the initial condition for our first simulation from 0 to 400My. Right: distribution of the particles surviving at 400 My, just before the planetary instability, which constitutes the initial condition for our second simulation, from 400My to 4.5Gy.  Particles are colored according to their lifetime in the simulation for which they represent the initial conditions. Thus, only red particles survive until the end of the simulation. 
Curved dashed lines in the top panels represent the boundaries for Earth-, Mars-, and Jupiter-crossing orbits (from left to right).  Vertical dashed lines show the main MMRs between asteroids and Jupiter.
\label{fig4}}
\end{figure*}

Figure \ref{fig4}, shows snapshots of the evolution of the Asteroid
Belt in our simulations, starting from the final 2-planet Grand Tack configuration.  The left panel illustrates the initial
system, corresponding to the solid black lines of Fig. \ref{fig3}. 
Considering the evolution described in section \ref{method} (with
planets on pre-instability orbits), and using Case 1 (low-eccentricity
giant planets orbits from \citet{Morbidelli07}), after 400 My of evolution
we obtain the distribution depicted in the right
panel of Figure \ref{fig4}. The colors of the particles in the left panels correspond to the particles' lifetimes in this simulation.  Results related to Case 2 (giant
planets with more eccentric orbits), will be considered only for
measuring the influence of the primordial eccentricities of Jupiter and
Saturn  and presented in section \ref{ejs}. 

 The distribution after 400~My is already closer to the current one
 (see Fig. \ref{fig2}) because the very high eccentricity objects have been removed
 due to interactions with the planets whose orbits they
 crossed. The main difference with the current main belt distribution is that our simulated belt extends inwards of 2.2~AU in semimajor axis. This is because the powerful $\nu_6$ resonance is not in place yet, given that Jupiter and Saturn are still in resonance and on quasi-circular orbits. This extension of the main belt towards the terrestrial planets is precisely the {\it E-belt} advocated in \citet{Bottke12}. 

Although not shown, the distribution after 700 My of
 evolution is nearly indistinguishable from that shown in the right
 panel of Fig. \ref{fig4}. This demonstrates that the exact timing of
 the giant planet instability is not important for this study.

In order to better understand the evolution during the first 400 My,
by looking in more detail at Fig. \ref{fig4}\footnote{An animation of
  the entire evolution, from 0 to 4.5 Gy, can be found electronically
  at extranet.on.br/rodney/rogerio/asteroids.mp4}, we see that a very strong depletion ($\sim$ 71\%) has
occurred,
approximately half in the first 100~My. This is due to the large
collection of particles on clearly unstable orbits, as noted above. In
fact, this happened mostly to objects with $e>$ 0.4, due to either
collision with the terrestrial planets (10\% of the removed particles)
or Jupiter (1.5\%) or ejections onto hyperbolic orbits (88\% of the
removed particles).  Only 0.5\% of the removed particles collided with
the Sun. This number is much smaller than that for the current
Near-Earth asteroids \citep{Farinella94,Gladman97}. 
This is because the giant planets are on more circular orbits, which makes the mean motion resonances much less effective in rising eccentricities than in the current Solar System (mean motion resonances are stable if Jupiter is on a circular orbit and would not open gaps, \citet{MorbyBook}). Also, the orbits of the giant planets are closer to each other, so the $\nu_6$ secular resonance is further out in the belt. Consequently, neither the mean motion nor the secular resonances are able to lift the particles' eccentricities to unity.

Although the distribution of particles at the end of the Grand
Tack evolution covers a wide range of eccentricities, semimajor axes, and
inclinations  (as shown in the left panel of Fig. \ref{fig4}), in order
to estimate the Asteroid Belt mass \citet{Walsh11}
consider only particles in the ``belt region'', with 
$q>$ 1.9 AU and $a<$ 3.2 AU.  In this region, our depletion ratio is
$\sim$ 24\% during the first 400 My and $\sim$ 16\% within the first
100 My.  The Grand Tack model places $\sim 1.3 \times$ 10$^{-3}$
M$_{\oplus}$ of S-types asteroids in the ``belt region'', and three
times as much of C-types asteroids, which implies a total mass of
$\sim 5.2 \times$10$^{-3}$ M$_{\oplus}$ for the post-Grand Tack main
belt.

Thus, after 400 My, we still have $\sim$76\% of the mass remaining in
the primordial main belt, which is $\sim$3.9$\times$10$^{-3}$
M$_{\oplus}$, or about 6-7 the current mass of the current Asteroid Belt
($\sim$6$\times$10$^{-4}$ M$_{\oplus}$).

\subsection{After the planetary instability} \label{post-LHB}

The second phase of the numerical simulations begin after the
planetary instability. Here, as explained in section \ref{method}, the
planets are instantaneously placed onto their current orbits. With the
new orbits, the secular resonances (particularly the $\nu_6$ resonance
at 2.05~AU) appear with their full power and several mean motion
resonances with Jupiter become unstable due to the larger
eccentricities of Jupiter and Saturn. The lifetimes of the particles that constituted the initial conditions for this second simulation are illustrated via a color scale in the right panels of Fig.~\ref{fig4}. 

\begin{figure}[h!]
\includegraphics*{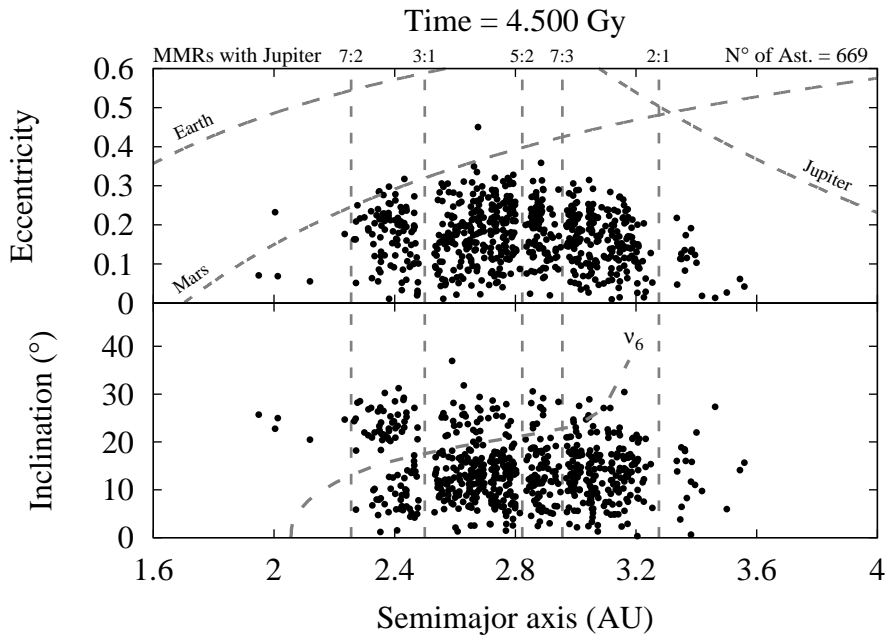}
\caption{Configuration of the system 4.1 Gy after the planetary instability (i.e. at the current time). 
Black dots represent the asteroids the survived in the simulation. Curved dashed lines in the top panel represent the boundaries for Earth-, Mars-, and Jupiter-crossing orbits (from left to right). In the bottom panel, the curved dashed line represents the current location of the $\nu_6$ secular resonance. Vertical dashed lines show MMRs between asteroids and Jupiter.
 \label{fig5}}
\end{figure}

Figure \ref{fig5} shows the final orbital distribution of the
particles surviving at the end of the full simulation, at 4.5
Gy. 
One may note that there is a nice qualitative
match between the current Asteroid Belt distribution and that produced
in the simulation (compare with Fig. \ref{fig2}). The range of distribution in semimajor axes,
eccentricities, inclinations, and the structure of the Kirkwood and
secular resonant gaps look very similar in the two cases.

\begin{figure*}[h!]
\begin{subfigmatrix}{2}
\includegraphics*{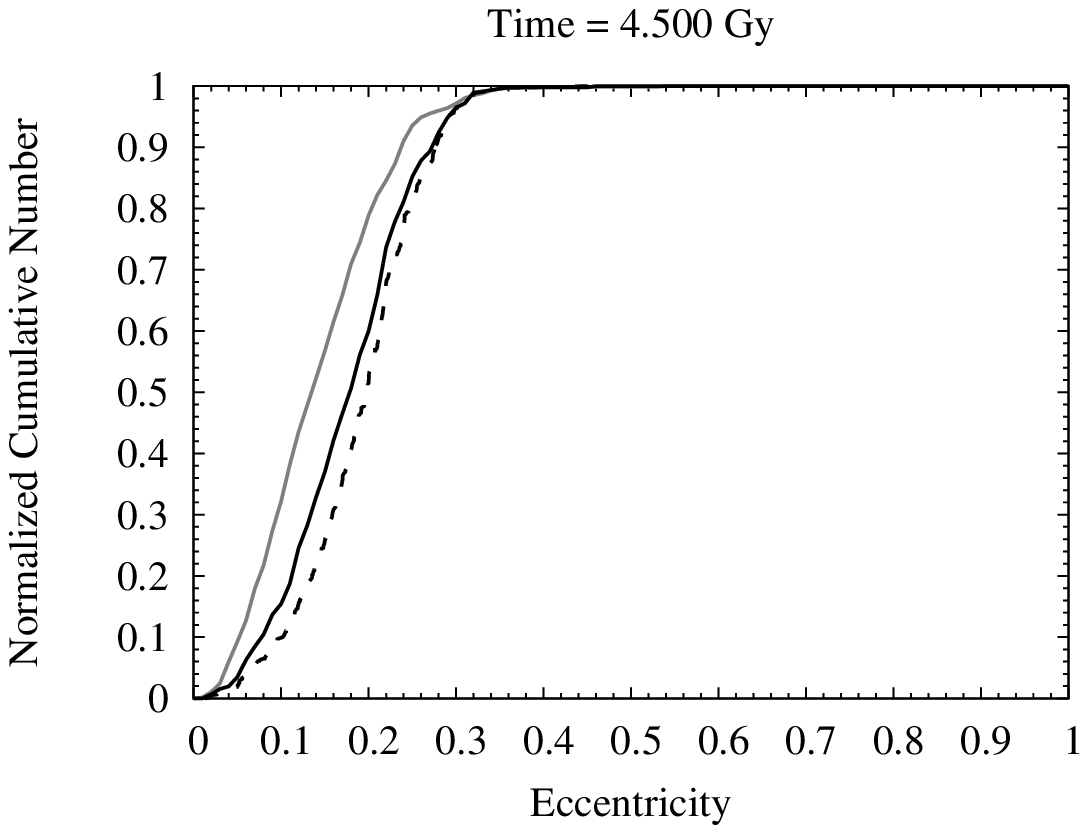}
\includegraphics*{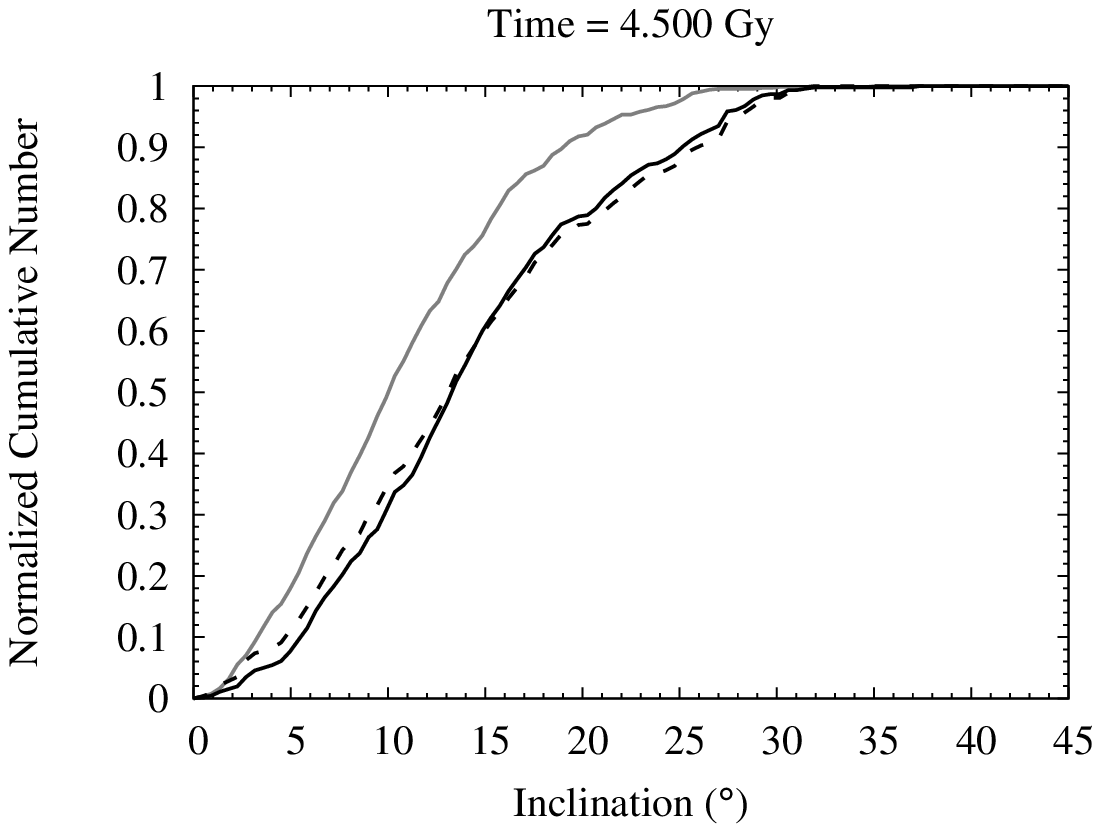}
\includegraphics*{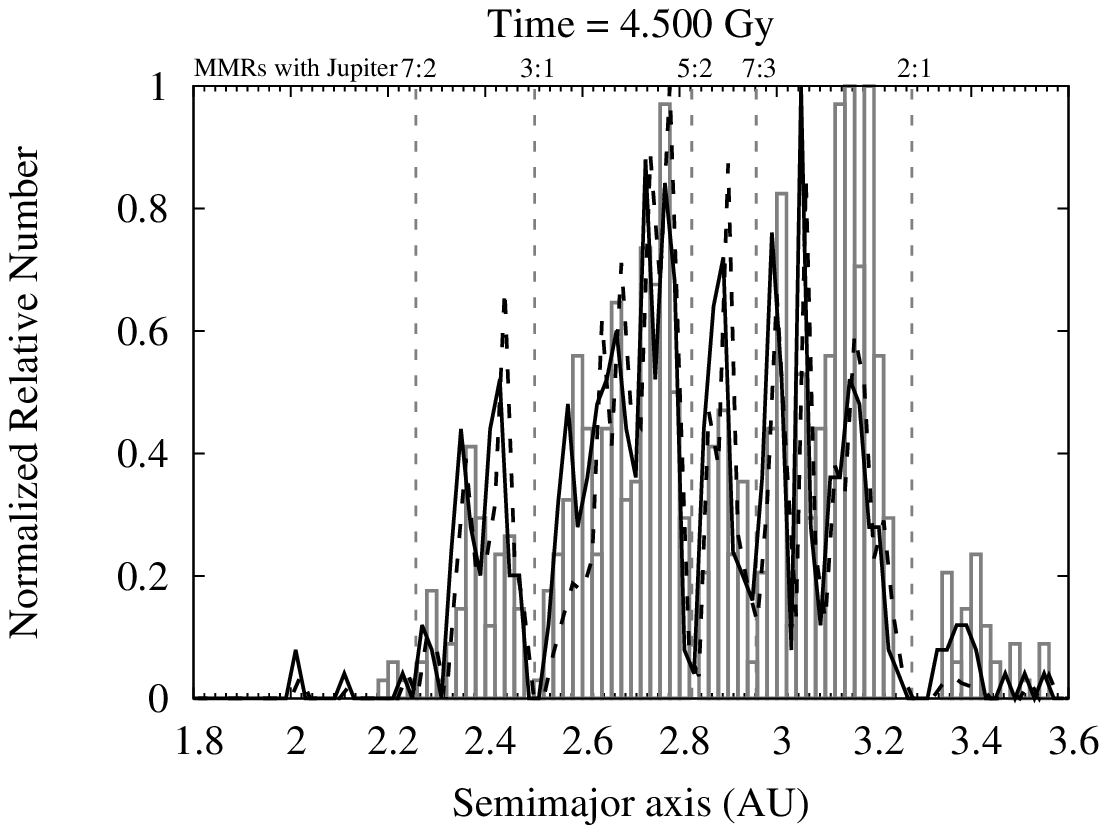}
\includegraphics*{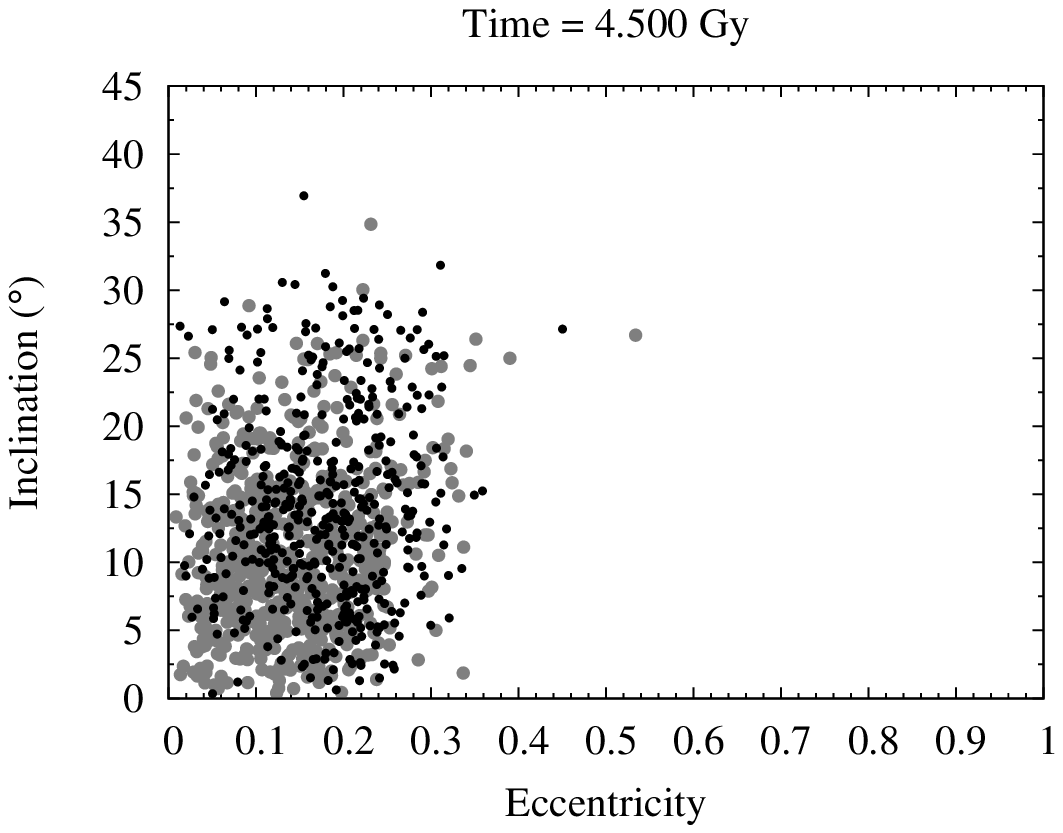}
 \end{subfigmatrix}
\caption{Normalized cumulative number of asteroids as a function of the eccentricity (top left) and inclination (top right). Normalized relative number of asteroids as a function of the semimajor axis (bottom left). The width $\Delta a$ of the bins of the histogram is 0.02 AU for the semimajor axis. In all these panels the gray line refers to the real distribution, the solid line shows the final distribution in our simulation and the dashed line shows the distribution re-weighted according to the probability that each integrated particle exists at the end of the 4-planet Grand Tack simulation. Bottom right: inclination as a function of the eccentricity. Large gray dots represent real asteroids with H$<$10 from the Minor Planet Center catalog. Small black dots represent the objects surviving after 4.5 Gy in our simulation.
\label{fig6}}
\end{figure*}

Fig. \ref{fig6} compares the final
distribution of the surviving simulated particles with the current Asteroid Belt orbital distribution, in a manner similar to what we did in Fig.~\ref{fig3} for the initial distribution.

Fig.~\ref{fig6} (bottom left panel) compares the final semimajor axis distributions. The visual comparison is very satisfactory, with the exception of the zone between the 7:3 and 2:1 MMRs, which is more relatively populated by real asteroids than in our model. To be more quantitative, the fraction of the total asteroid population orbiting between major Kirkwood gaps is reported in Table~\ref{ratios} for both the observed and the model populations. Again, we see a fairly good agreement.

\begin{table}
\centering
\begin{tabular}{lllll}
\hline
Kirkwood gap ratios & Zone 1 & Zone 2 & Zone 3  & Zone 4  \\
\hline
 Major Belt               &   0.106     &     0.373       &    0.093        &  0.427  \\
 Simulation              &   0.147     &     0.394       &    0.141        & 0.318 \\
\hline
\end{tabular}
\caption{Fraction of the asteroid population orbiting between major Kirkwood gaps. All fractions are caulculated dividing the number of asteroids in between two adjacent Kirkwood gaps by the total number of asteroids between the 7:2 and the 2:1 MMRs. Thus: Zone 1 represents the fraction of asteroids between the MMRs 7:2 and 3:1. Zone 2 the fraction between the 3:1 and 5:2 MMRs. Zone 3 between the 5:2 and 7:3 MMRs, and Zone 4 between the 7:3 and 2:1 MMRs.}
\label{ratios}
\end{table}

The inclination distribution (Fig. \ref{fig6} top right) does not
change very much (compare with Fig. \ref{fig3}) from the beginning to the end of our simulations,  moving just slightly towards larger inclination values. Indeed low inclination asteroids are removed somewhat more frequently than high inclination asteroids during the giant planet instability. Thus, our final inclination distribution is moderately more excited than the observed distribution. 

This is particularly apparent in Fig.~\ref{fig5}, in which we see (by comparison with the observed distribution in Fig.~\ref{fig2}) that the ratio between the number of asteroids above and below the $\nu_6$ resonance in the inner belt ($a<$ 2.5 AU) is much larger in our model ($\sim$1.2) than in reality ($\sim$0.09). However, if the initial inclination distribution (at the end of the Grand Tack) had been less excited, the final population ratio would have been much closer (for instance we obtain a population ratio of $\sim$0.1 if we truncate the initial distribution at 20 degrees). Thus, this is clear evidence that the inclination distribution at the end of the Grand Tack is too excited. This result may suggest that the Grand Tack evolution occurred faster than simulated in \citet{Walsh11}, or that the planets had not yet reached their current masses, or there was a stronger gas-drag effect.  
 
In the eccentricity panel (Fig. \ref{fig6} top left), despite some
differences that are still observed between the real and the model
distributions, the match appear reasonably good, particularly if one
considers how different the Grand Tack eccentricity distributions were
originally (compare with Fig. \ref{fig3}).  This shows that the Grand
Tack model can be consistent with the current eccentricity
distribution. 

\begin{figure*}[h!]
\begin{subfigmatrix}{2}
\includegraphics*{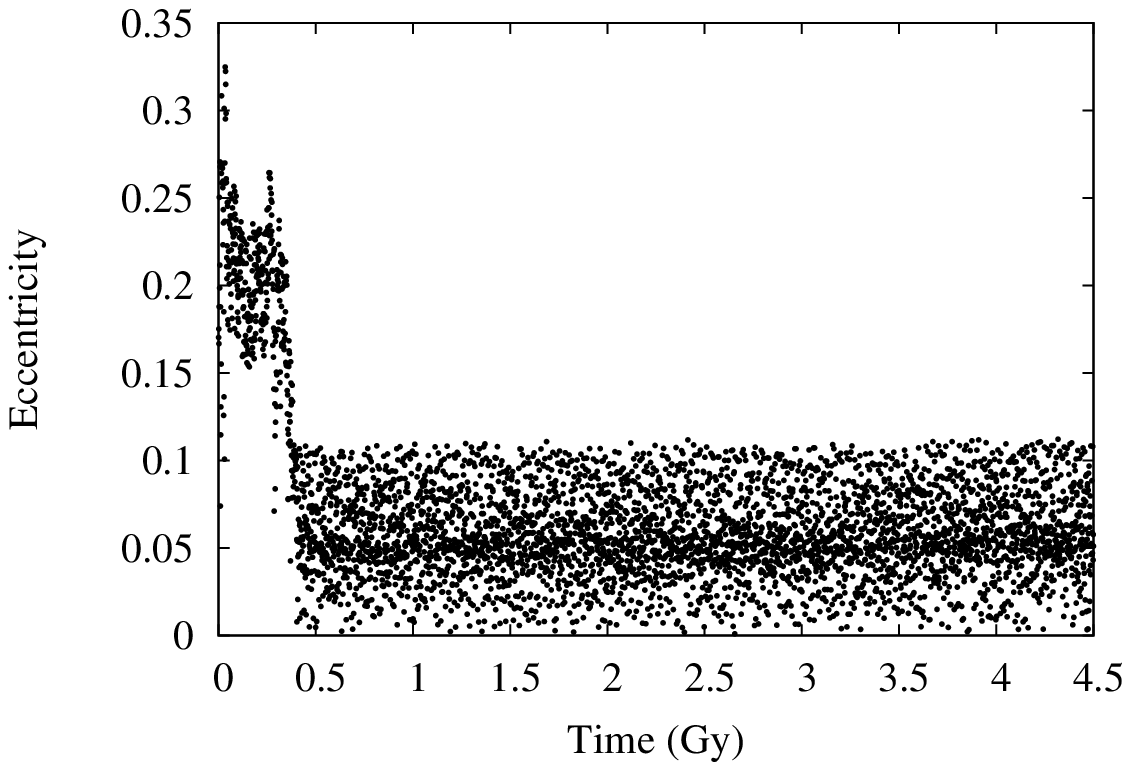}
\includegraphics*{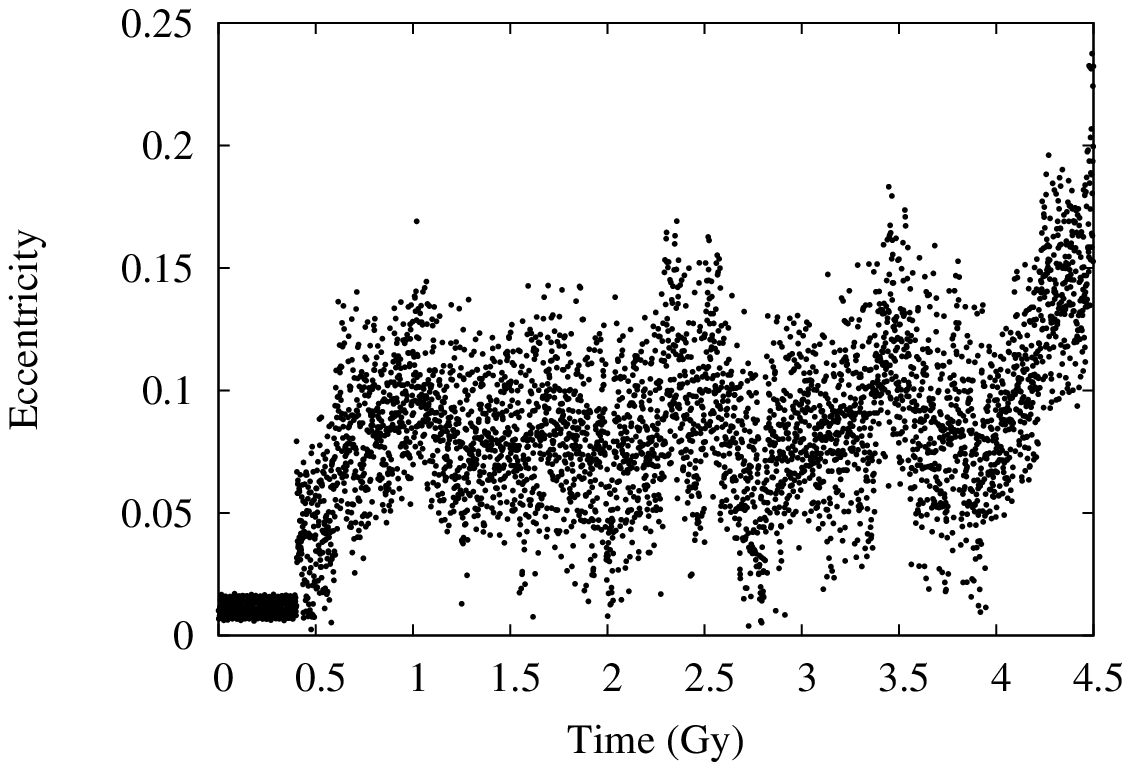}
\includegraphics*{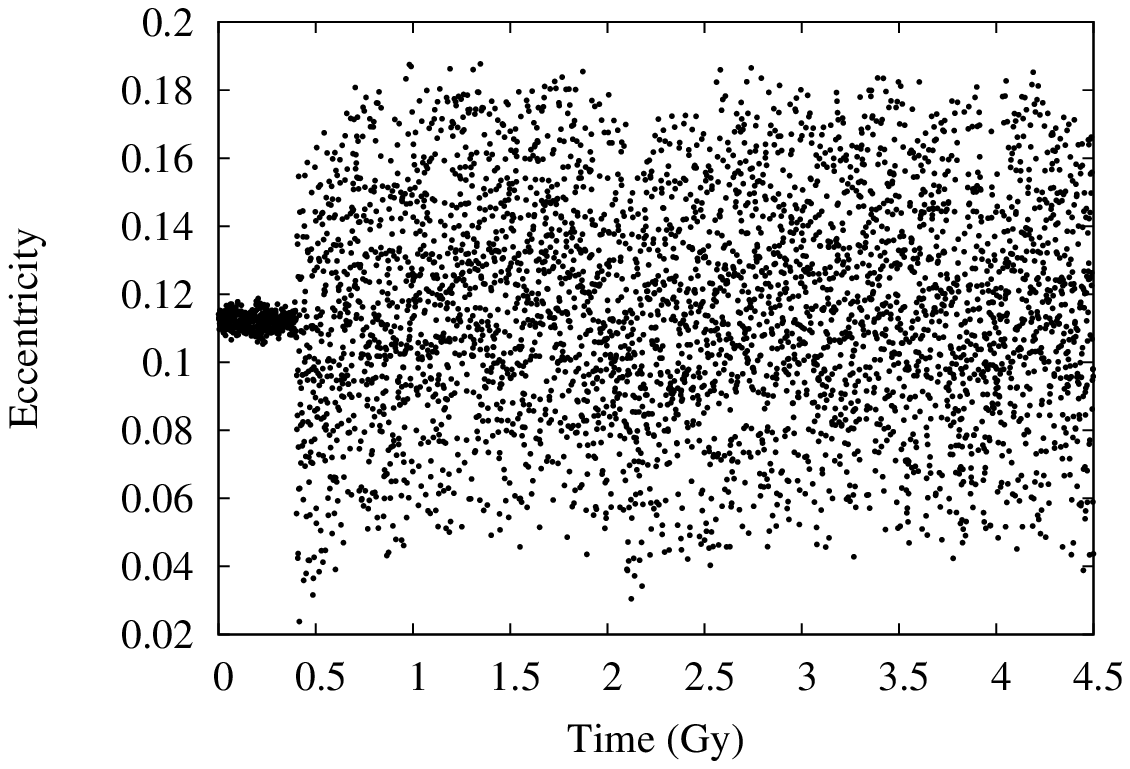}
 \end{subfigmatrix}
\caption{Evolution of the orbital eccentricity of three asteroids (test particles) during 4.5 Gy. Top left panel ilustrates a case where the eccentricity of the asteroid decreases after the LHB (at 0.4 Gy). Top right panel shows the opposite case, where the eccentricity of the asteroid increases. Bottom panel represents the majority of the cases, where only changes $<$ 0.05 occur in the mean eccentricity value.
 \label{fig7}}
\end{figure*}

The reason for which the final eccentricity distribution is much
  more in agreement with the observed distribution than the initial
  one is twofold. First, as explained above, large eccentricity
  asteroids tend to be removed because they are unstable on a
  timescale shorter than the Solar System lifetime. Second, when
Jupiter transitions form a quasi-circular to an eccentric orbit, the
asteroid's secular motion sees a value of the forced eccentricity
different from before (which was almost zero). Depending on the values
of the difference in longitude of perihelia between the asteroid and
Jupiter when this happens, some asteroids will find themselves at the
minimum of the new secular eccentricity cycle (and, possibly, by
achieving a larger eccentricity they will be removed by encounters
with the terrestrial planets) while others will be at the maximum of
the secular eccentricity cycle. Others, of course, will be at
intermediate values. By this mechanism, the mean eccentricity of some
asteroids can either decrease or increase during the jump in Jupiter's
eccentricity (Fig. \ref{fig7}). \citet{Minton11} discuss an analog
process in the case of a secular resonance sweeping through an excited
belt.  Specifically, in our simulations we observed that $\sim$77\% of the survivor asteroids kept their mean eccentricity
nearly unchanged ($\Delta$$\bar{e}$ $<$ 0.05) when passing through the
dynamical instability (Fig. \ref{fig7}, bottom
panel). On the other hand, within those $\sim$23\% that have had
their mean eccentricity changed in more than 0.05 in this phase, the ratio decrease/increase in 
mean eccentricity values was about half ($\sim$51\%
decrease and $\sim$49\% increase -- Fig. \ref{fig7}, top left
and right panels, respectively). We considered the first and last 100
My of simulation to compute and compare the mean values of
eccentricities.

This reshuffling of eccentricities explains why the solid and dashed black curves in the top left panel of Fig.~\ref{fig6} are so similar to each other. Remember (see sect.~\ref{method}) that the dashed distribution is obtained from our nominal final distribution (solid curve) by weighting each integrated particle according to its probability to exist at the end of the 4-planet Grand Tack simulation. Because the distribution at the end of the 4-planet Grand Tack simulation is more skewed to large values than that in the 2-planet case (see Fig.~\ref{fig3}), the particles in our integration with an initial low eccentricity have a weight well smaller than 1, while those with an initial large eccentricity have a weight larger than 1. Thus, if the evolution just following the Grand Tack had just preserved low-eccentricity asteroids, the final solid and dashed distribution would have been very different. Instead, the fact that a significant number of initially large-$e$ particles acquire a small eccentricity erases the initial difference in eccentricity distributions. This result is important because it shows that the evolution subsequent to the Grand Tack generates a final eccentricity distribution of asteroids that is quite insensitive of the initial one and in broad agreement with the observed one.

Concerning the mass depletion of the asteroid belt, we find that 
during the second phase of the simulation, from 400 My to 4.5 Gy, there
was a depletion of $\sim$77\% of the particles. This depletion is
consistent with that observed in the simulations of
\citet{Morbidelli10} of the instability phase (about 50\% depletion),
followed by the additional 50\% depletion observed in \citet{Minton10}
due to asteroids in diffusive regions, unstable on a longer timescale.

In terms of mass, we need again to consider the ``belt region''. In this region, the depletion from 400
My to 4.5 Gy was $\sim$58\%. Recall that after the first 400 My,
$\sim$76\% of the particles survived in the ``belt region'' (representing
$\sim$3.9$\times$10$^{-3}$ M$_{\oplus}$). Thus, additionally,
$\sim$2.3$\times$10$^{-3}$ M$_{\oplus}$ was lost during this
phase. Therefore, the final estimate for the surviving mass in the
Asteroid Belt in our simulation is $\sim$1.6$\times$10$^{-3}$
M$_{\oplus}$, which represents about 2-3 the current mass. Given the
uncertainties in the implementations of the Grand Tack model (see the
Supplementary Information in \citet{Walsh11} for information on the
wide range of test cases explored), we consider this to be a
satisfactory match of the current Asteroid Belt mass.

Lastly, although there is overall consistency between our results and
the current main belt, our results (bottom right panel of Fig.~\ref{fig6}) show a lack of particles with low
inclination at low eccentricity. In order to improve the results, either the Grand Tack simulations should have produced more particles in this region, or during the instability phase Jupiter should have had acquired a temporary orbit with a larger eccentricity than the current one, enhancing the reshuffling of the eccentricity distribution discussed above. The last option is actually likely to have taken place, because simulations of the instability \citep{Nesvorny12} show that the eccentricity of Jupiter's orbit is partially damped afterwards, which implies that it had to be larger in the past. By substituting the primordial Jupiter's orbit with the current orbit, we clearly missed this phase.

Our results are complementary to those of \citet{Roig15}. In
that paper the authors considered the dynamics of the giant planet
instability from the most successful simulations of \citet{Nesvorny12}
but did not include the presence of the terrestrial planets. They
concluded that in order to reproduce the present Main Belt, the
primordial belt should have had a distribution peaked at
$\sim10^{\circ}$ in inclination and at $\sim0.1$ in eccentricity. Here
we start from the actual asteroid distribution at the end of the Grand
Tack model, include the effects of terrestrial planets, but simplify
the dynamical instability phase. We find that the Grand Tack
inclination distribution is indeed peaked at 10 degrees and the eccentricity
distribution, once reshaped mostly due to scattering by the
terrestrial planets, rapidly evolves to one that peaks at about 0.1-0.2. Hence, our final distribution is in
good agreement with the observed one, with a sight deficit of
asteroids in orbits with low eccentricity and a slight overabundance of large inclination bodies.
  
\subsection{Primordial eccentricity of Jupiter and Saturn} \label{ejs}

Up to this point we have only considered pre-instability orbits of the
giant planets as shown in Case 1 (Fig. \ref{fig1} left, section
\ref{method}). Here we explore the putative case of more eccentric
pre-instability giant planet orbits, Case 2 (Fig. \ref{fig1} right,
section \ref{method}), and compare with the results of Case 1
presented above.

\begin{figure*}[h!]
\begin{subfigmatrix}{2}
\includegraphics*{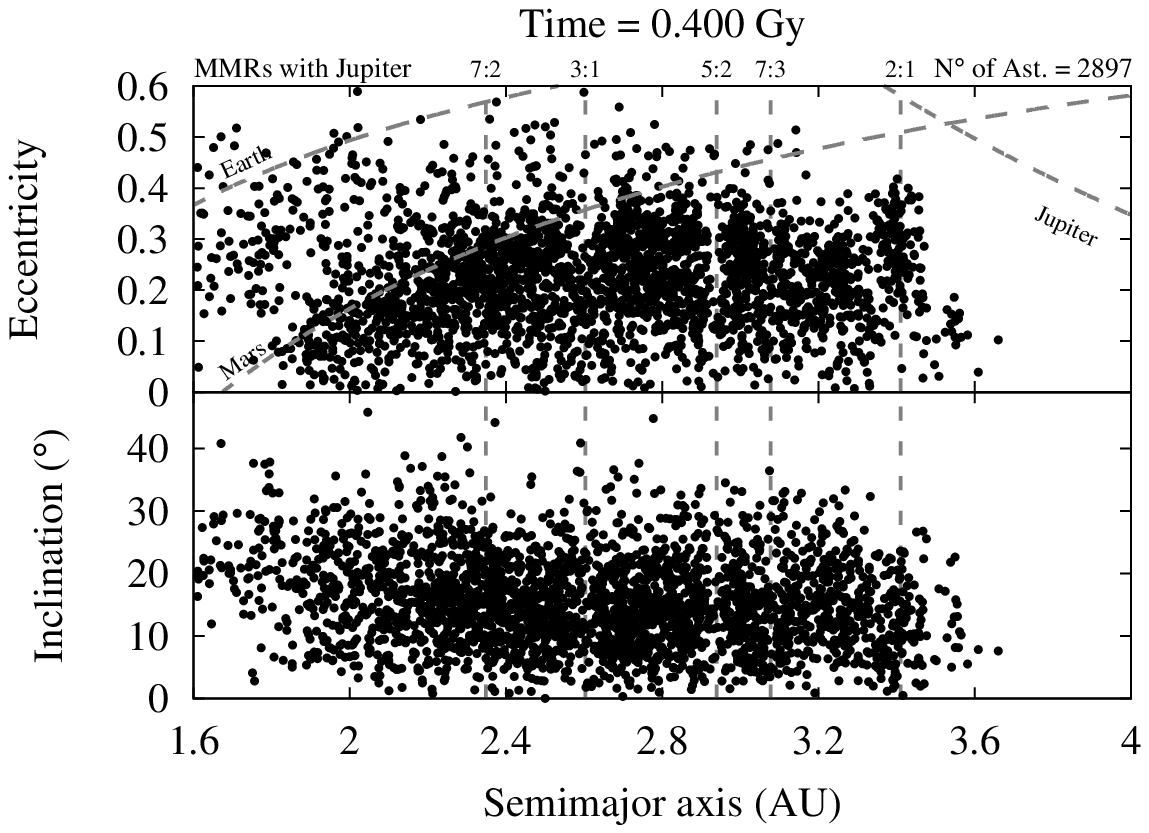}
\includegraphics*{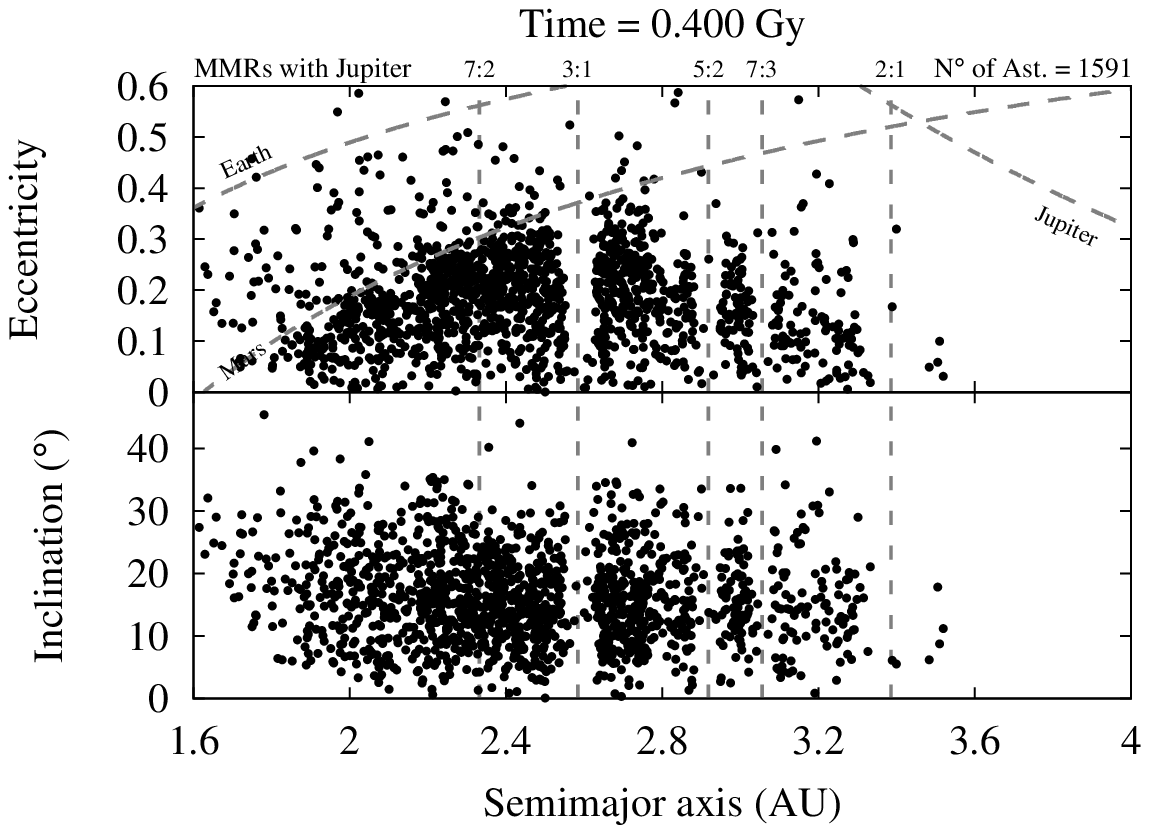}
\includegraphics*{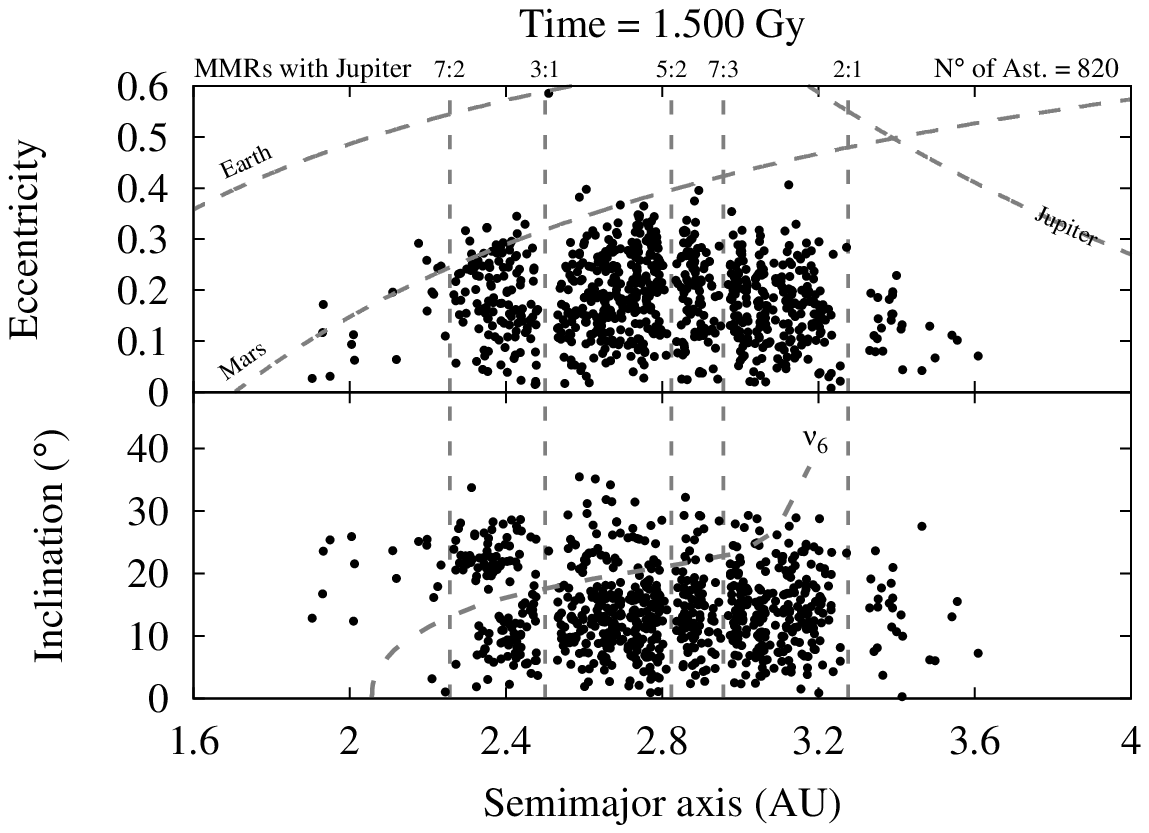}
\includegraphics*{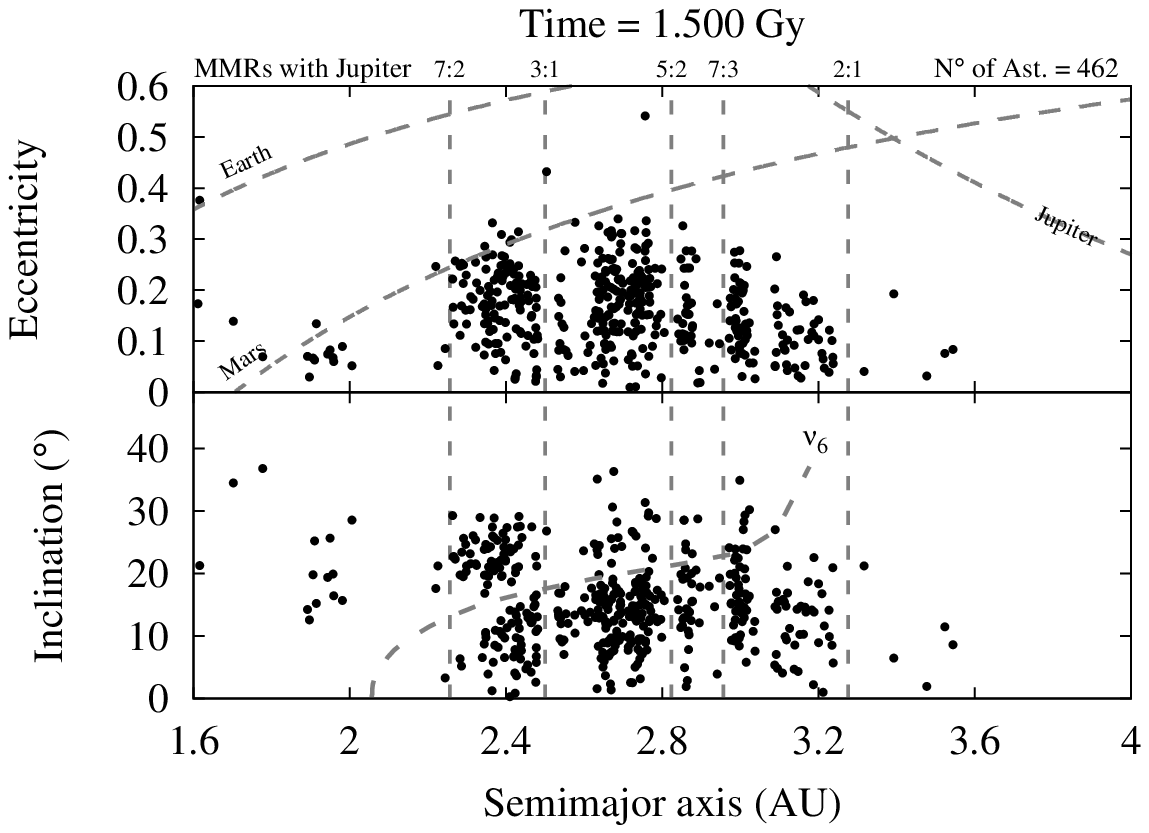}
 \end{subfigmatrix}
\caption{Snapshots of the evolution comparing the results obtained in the simulations of Case 1 (circular Jupiter and Saturn) and Case 2 (eccentric Jupiter and Saturn).
Left column: Case 1. Right column: Case 2. The evolutions of Jupiter and Saturn in both cases, before the planetary instability, are shown in Fig. \ref{fig1}. Top: Before the planetary instability. Bottom: after the planetary instability. 
Curved dashed lines in the top of each panel represent the boundaries for Earth-, Mars-, and Jupiter-crossing orbits (from left to right). In the bottom of the bottom panels, the curved dashed line represents the current location of the $\nu_6$ secular resonance.
Vertical dashed lines show MMRs between asteroids and Jupiter.
 \label{fig8}}
\end{figure*}

\begin{figure*}[h!]
\begin{subfigmatrix}{2}
\includegraphics*{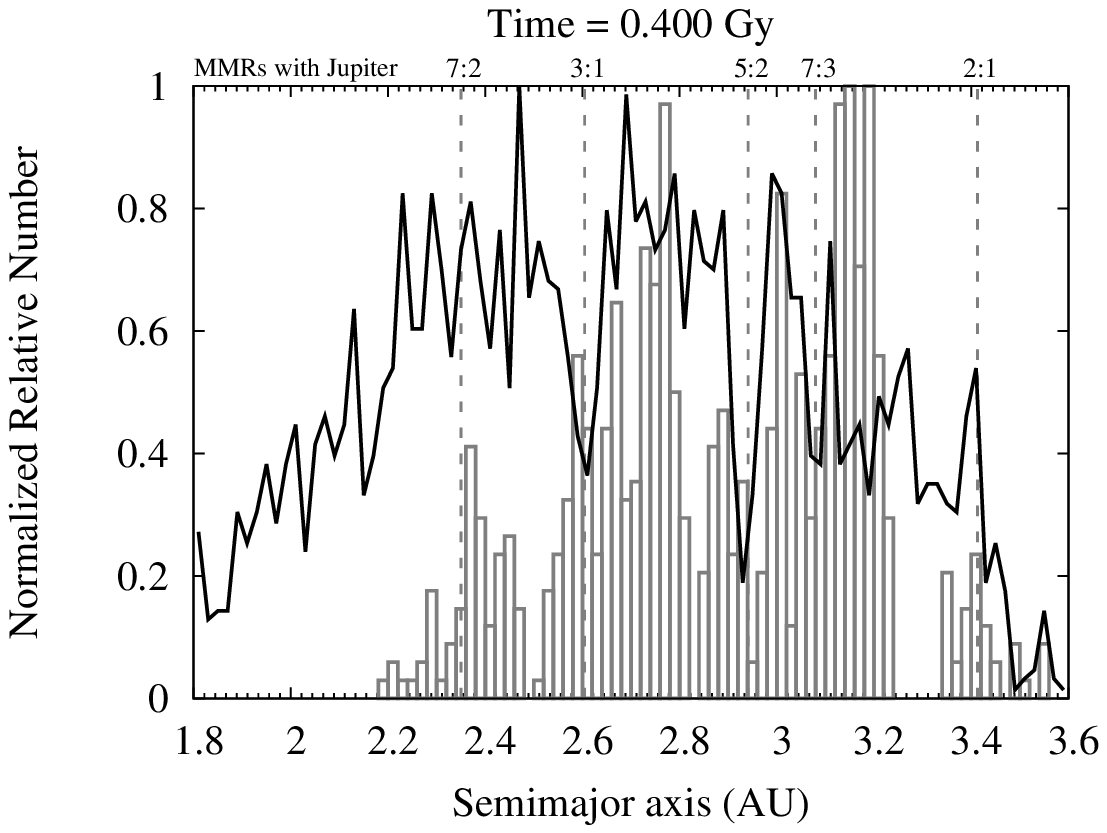}
\includegraphics*{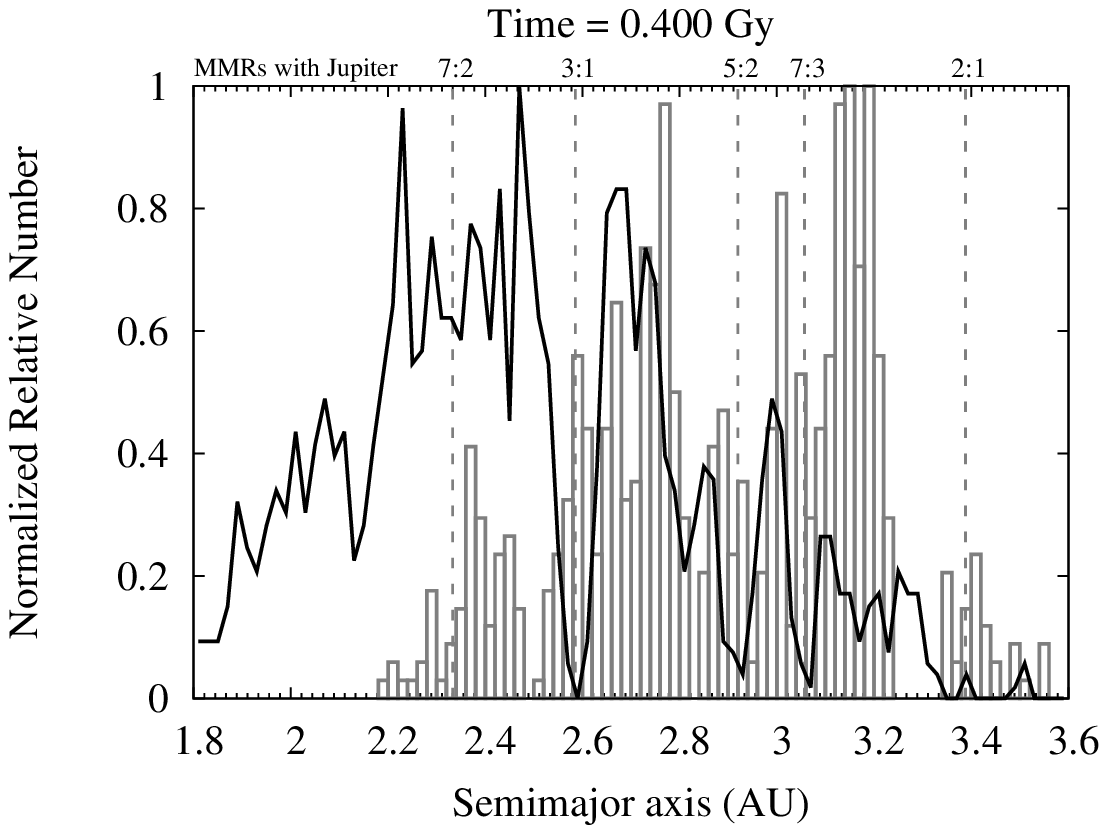}
\includegraphics*{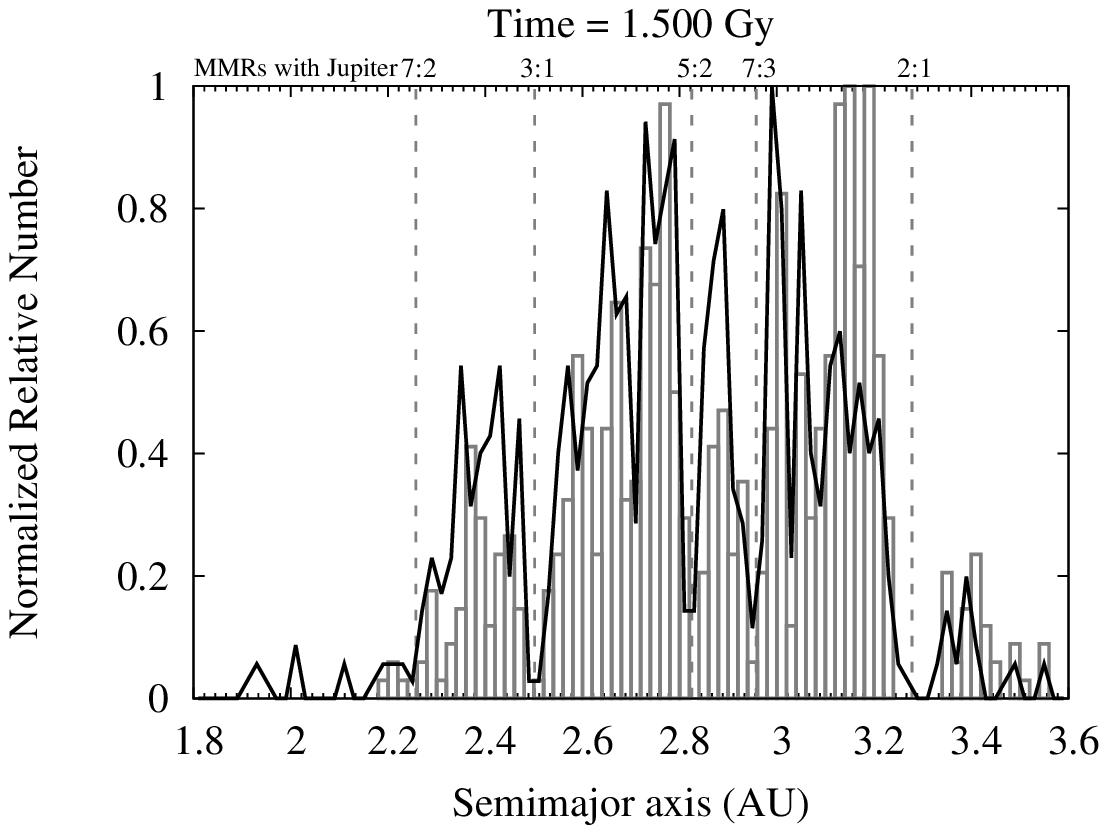}
\includegraphics*{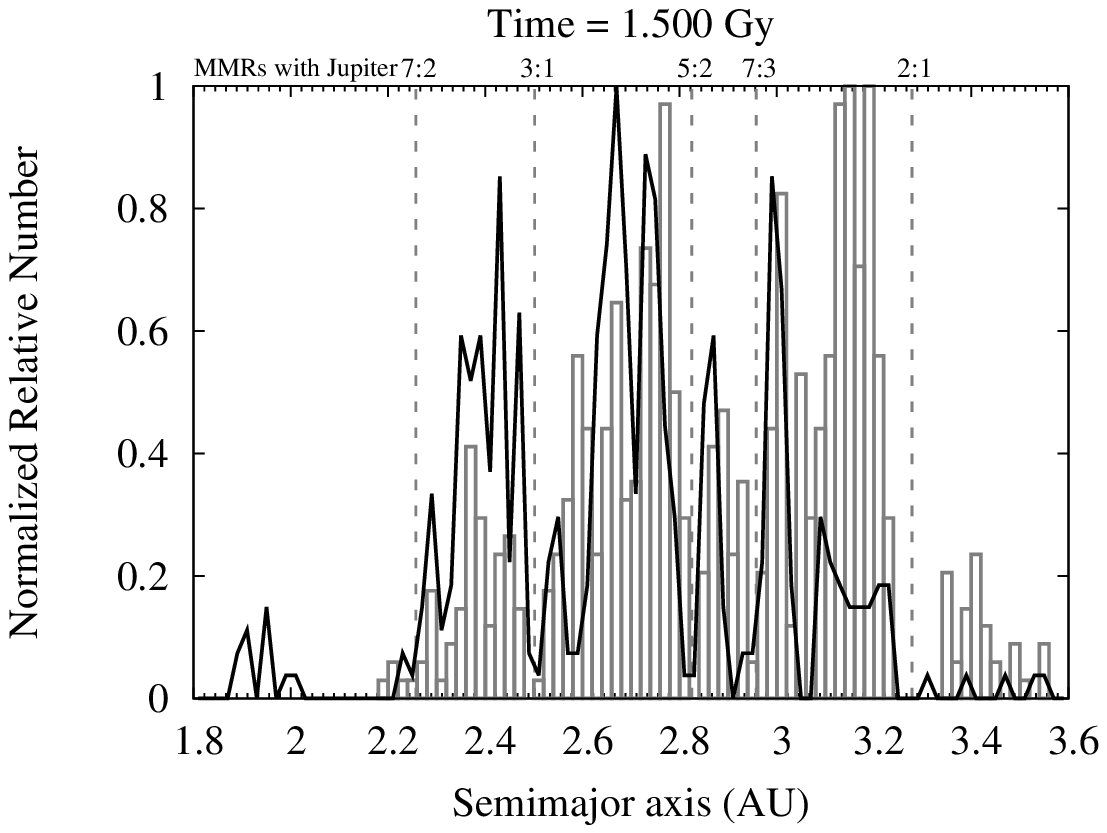}
 \end{subfigmatrix}
\caption{Comparison of the semimajor axis distributions obtained at the end of the simulations of Case 1 (circular Jupiter and Saturn, left panels) and Case 2 (eccentric Jupiter and Saturn, right panels). Top: asteroid distribution before the planetary instability. Bottom: after the planetary instability. 
\label{fig9}}
\end{figure*}

The comparison between the results of the simulations in Case 1 and
Case 2 is shown with snapshots of the evolution in Fig. \ref{fig8}
(with Case 1 in the left column and Case 2 in the right column) and
final distribution histograms in Fig. \ref{fig9}. The distributions are shown
both before and after the planetary instability phases (up to 1.5
Gy\footnote{we did not extend the simulation to 4.5 Gy because after
  1.5 Gy the structure of the Asteroid Belt does not change much over
  time, as one may see by comparing Figs. \ref{fig6} and \ref{fig9} (bottom left panel), as well as by the analysis of the animation found electronically at extranet.on.br/rodney/rogerio/asteroids.mp4}). Case 2 results in a
larger depletion. After 0.4 Gy Case 1 lost 72\%, and Case 2 84\% of
the primordial asteroids.

More importantly, Case 2 shows a wide gap just outside of the 3:1 MMR
at 2.5 AU. This gap is not visible in the simulation of Case 1 and it
does not exist in today's Asteroid Belt.  Such a gap corresponds to
the location of the 3:1 mean motion resonance with Jupiter, when
Jupiter was on the pre-instability orbit at $a=$ 5.4~AU. With the
moderate eccentricity orbits of the giant planets considered in Case
2, the resonance is unstable and opens a gap in the asteroid
distribution in 400~My.  Assuming that the instability happened
late, so that there is enough time to open the gap at the 3:1 MMR and
that there was no interloper planet that would cause scattering of the
asteroids' semimajor axes, this gap becomes fossilized and is never
refilled in the subsequent, post-instability evolution. The same is
true for other resonances with Jupiter (5:2 MMR at $\sim$2.9 AU, 7:3
MMR at $\sim$3.1 AU, 2:1 MMR at 3.4 AU).  Instead, in Case 1, where
the eccentricities of the giant planets are much smaller, these
resonances are far less powerful and do not open gaps. This result is
consistent with that in \cite{Morbidelli10} and constrains the
eccentricities of Jupiter and Saturn characterizing their
pre-instability orbits, i.e., the primordial eccentricities of Jupiter
and Saturn were likely quite low (as in Case 1 simulation).

\newpage
\section{Conclusions} \label{conclusion}

In this paper we evaluated the evolution of the primordial Asteroid
Belt resulting from the Grand Tack model throughout the entire age of the Solar
System. We assumed that the configuration just after the end of
the Grand Tack phase is the one that emerges when the gas nebula
dissipates and only solid material remains in the Asteroid Belt and
planetary regions.  The Grand Tack model can explain the mixing of the
taxonomic classes of the bodies within the asteroidal region. However,
as a major drawback, it predicts a primordial Asteroid Belt that seems 
too excited when compared to the real
Asteroid Belt.

In order to see how the subsequent evolution of the Solar System would
affect the excited primordial belt, and if the Grand Tack model
predictions are coherent with the current Asteroid Belt, we considered
two phases of evolution:

\begin{itemize}

\item The first dates back to the first 400 My, before the planetary
  instability, but after the terrestrial planets have been fully
  formed in orbits more planar and circular than today
  \citep{Brasser13}. Also in this phase we considered Jupiter and
  Saturn locked in their 3:2 MMR with Jupiter initially at $a$ = 5.4
  AU, with both planets in quasi planar and circular orbits as
  well. During this phase we noticed a huge depletion of the asteroid population
  (about $\sim$71\%), mostly due to ejection ($\sim$88\%). Most of the removed asteroids had initially 
  $e>$ 0.4. Thus, in the ``belt region'' ($q>1.9$~AU and $a<3.2$~AU) only about 
  $\sim$24\% of the initial mass was lost,
  leaving $\sim$3.9$\times$10$^{-3} M_{\oplus}$ (6--7 the current mass
  of the current Asteroid Belt).

\item The second phase starts just after the end of the planetary
  instability. We simulated the planetary instability as a
  discontinuous event, by instantaneously moving all planets to their
  current orbits. In other words, as explained in section
  \ref{method}, we considered that the planetary instability happened
  fast enough that the actual dynamical path that the planets followed
  between the transition from the pre-instability orbits to the
  current orbits had no important effects. Thus, by using the
  current planetary system, the locations and strengths of all
  resonances match exactly the real ones, and we could simulate the
  response of the Asteroid Belt to the change of the planetary
  orbits. This phase was carried from 400 My to 4.5 Gy and as a major
  result, it substantially changed the eccentricity distribution that,
  at the end, mostly resembles the current Asteroid Belt.

\end{itemize}

During the second phase, when the MMRs with Jupiter are unstable, and
the $\nu_6$ destabilizes the extended part of the Asteroid Belt
(E-Belt), another period of strong depletion occurs ($\sim$77\% of the bodies remaining from phase 1 were lost). Again, most of
the depletion  is due
to ejections and concerns bodies not in the ``belt region''. In terms of the remaining mass, the ``belt region''
had an additional $\sim$58\% of loss, ending with
$\sim$1.6$\times$10$^{-3} M_{\oplus}$ (2--3 the current mass of the
current Asteroid Belt), which is a satisfactory match of the current
Asteroid Belt's mass given the uncertainties in the implementations of
the Grand Tack model.

As for the orbital structure of the final Asteroid Belt, even
considering that we have a deficit of objects with low eccentricity with
low inclinations in our final distribution, the results are quite
good.  In fact, the inclination distribution evolves slightly towards a more excited distribution and is in the end 
somewhat too excited relative to the observed one. We infer that the inclination distribution at the end of the Grand Tack phase should have extended only up to $\sim 20^\circ$, instead of $\sim 30^\circ$ as in \citet{Walsh11}. 
The semimajor
axis distribution evolves from an almost uniform distribution to one
quite similiar to that of the real Asteroid Belt, not only in terms of
location of Kirkwood gaps but also in terms of the relative number of objects in each sub-belt 
region. The eccentricity distribution produced by the Grand Tack, initially skewed to large values, evolves to one similar to the observed
distribution once we take into account the post-Grand Tack dynamics.
Remarkably, this is the case both starting from the final distributions in the 2-planet and 4-planet Grand Tack simulations, despite these distributions being quite different from each other. 
The fact that \citet{Roig15}, with a very different approach, found a
similar result gives support to the claim that the Grand Tack scenario
is compatible with the current properties of the Asteroid Belt.

Finally, assuming that the instability happened late and that there
was no interloper planet that would disperse the semimajor axes of the
asteroids during the planetary instability, we show that primordial
eccentric orbits for Jupiter and Saturn would have produced fossil
Kirkwood gaps at misplaced positions that are not
observed. Thus, our results support the idea that the orbits of the
giant planets remained quasi-circular until the time of the instability.

\section*{Acknowledgments}

The authors would like to thank David P. O'Brien and Nathan Kaib for the detailed review, comments, and suggestions made to improve the manuscript.
R.D. acknowledges support provided by grants \# 2014/02013-5 and \#2015/18682-6, S\~ao Paulo Research Foundation (FAPESP) and CAPES. 
R. S. G. was supported by CNPq (grant \# 307009/2014-9). 
K. J. W. acknowledges support through I-SET, a NASA SSERVI program at SwRI. 
A. M.  acknowledges support from ANR project number ANR-13\{13-BS05-
0003-01 projet MOJO.
D. N. was supported by NASA's Outer Planet Research program.





\end{document}